\documentclass[aps,prl,twocolumn,superscriptaddress,groupedaddress]{revtex4-2}
\usepackage{graphicx,epstopdf} 
\usepackage{amsmath}
\usepackage{color}
\usepackage{float}
\usepackage{soul}
\usepackage{hyperref}
\usepackage{cleveref}
\usepackage{lipsum, babel}

\newcommand{\be}{\begin{equation}}
\newcommand{\ee}{\end{equation}}
\newcommand{\bea}{\begin{eqnarray}}
\newcommand{\eea}{\end{eqnarray}}
\newcommand{\bse}{\begin{subequations}}
\newcommand{\ese}{\end{subequations}}

\usepackage{multirow}
\begin{document}
\title{ Tunable room temperature magnetic skyrmions in centrosymmetric kagome magnet Mn$ _4 $Ga$_2 $Sn.}
\author{Dola Chakrabartty}
\affiliation{School of Physical Sciences, National Institute of Science Education and Research, HBNI, Jatni-752050, India}
\author{Sk Jamaluddin}
\affiliation{School of Physical Sciences, National Institute of Science Education and Research, HBNI, Jatni-752050, India}
\author{Subhendu K. Manna}
\affiliation{School of Physical Sciences, National Institute of Science Education and Research, HBNI, Jatni-752050, India}
\author{Ajaya K. Nayak}
\email{ajaya@niser.ac.in}
\affiliation{School of Physical Sciences, National Institute of Science Education and Research, HBNI, Jatni-752050, India}
 \date{\today}

\begin{abstract}
The successful realization of skyrmion-based spintronic devices depends on the easy manipulation of underlying magnetic interactions in the skyrmion-hosting materials. Although the mechanism of skyrmion formation in non-centrosymmetric magnets is comprehensively established, the stabilization process of different skyrmion-like magnetic textures in centrosymmetric magnets needs further investigation. Here, we utilize Lorentz transmission electron microscopy study to report the finding of a tunable skyrmion lattice up to room temperature in a centrosymmetric kagome ferromagnet Mn$ _4 $Ga$ _2 $Sn. We demonstrate that a controlled switching between the topological skyrmions and non-topological type-II magnetic bubbles can be realized at the optimal magnetic anisotropy. We find that the topological skyrmions are the energetically most stable magnetic objects in the centrosymmetric hexagonal magnets, whereas application of in-plane magnetic field stabilizes type-II magnetic bubbles as an excited state. The present study is an important step towards understanding of the skyrmion stabilization mechanism in centrosymmetric materials for their future applications.
\end{abstract}


\maketitle
\section{INTRODUCTION}

Magnetic skyrmions are topologically protected nontrivial chiral spin configurations that can move in a small cut-off current by avoiding defects in their path \cite{MnSi ultralow current,emergent electrodynamics,fege ultralow current}, and hence are considered as excellent candidates for the future high density based racetrack memory devices \cite{Tokura_review}. Skyrmions have been mostly observed in the non-centrosymmetric chiral magnets \cite{MnSi,FeCoSi,cu2seo3,CoZnMn,GaV4s8} and multilayer thin films \cite{blowing sk bubble, Ir/Fe/Co/Pd, Pt/Co}, where the competition between the Dzyaloshinskii–Moriya interaction (DMI) and the exchange interaction plays a significant role in stabilizing the underlying spin textures. In recent times, skyrmion-like whirling spin textures with various topological numbers are found in certain centrosymmetric magnets with uniaxial magnetocrystalline anisotropy (UMA) \cite{LSFMO,LSMO0P315,MnNiGa,Fe3Sn2,LSMO0P175,MnPdGa,NdCo5}. In these magnets, skyrmion-like spin textures can be obtained as a result of competing dipolar energy and the magnetic anisotropy. As the dipolar energy is one of the most important energy contributions  towards the  formation of skyrmions in centrosymmetric magnets,  an easy control over the shape, size, chirality as well as the topological charge of the skyrmions can be achieved  by tuning the magnetization and thickness of the sample.

\begin{figure*}
	\begin{center}
		\includegraphics[width= 15cm]{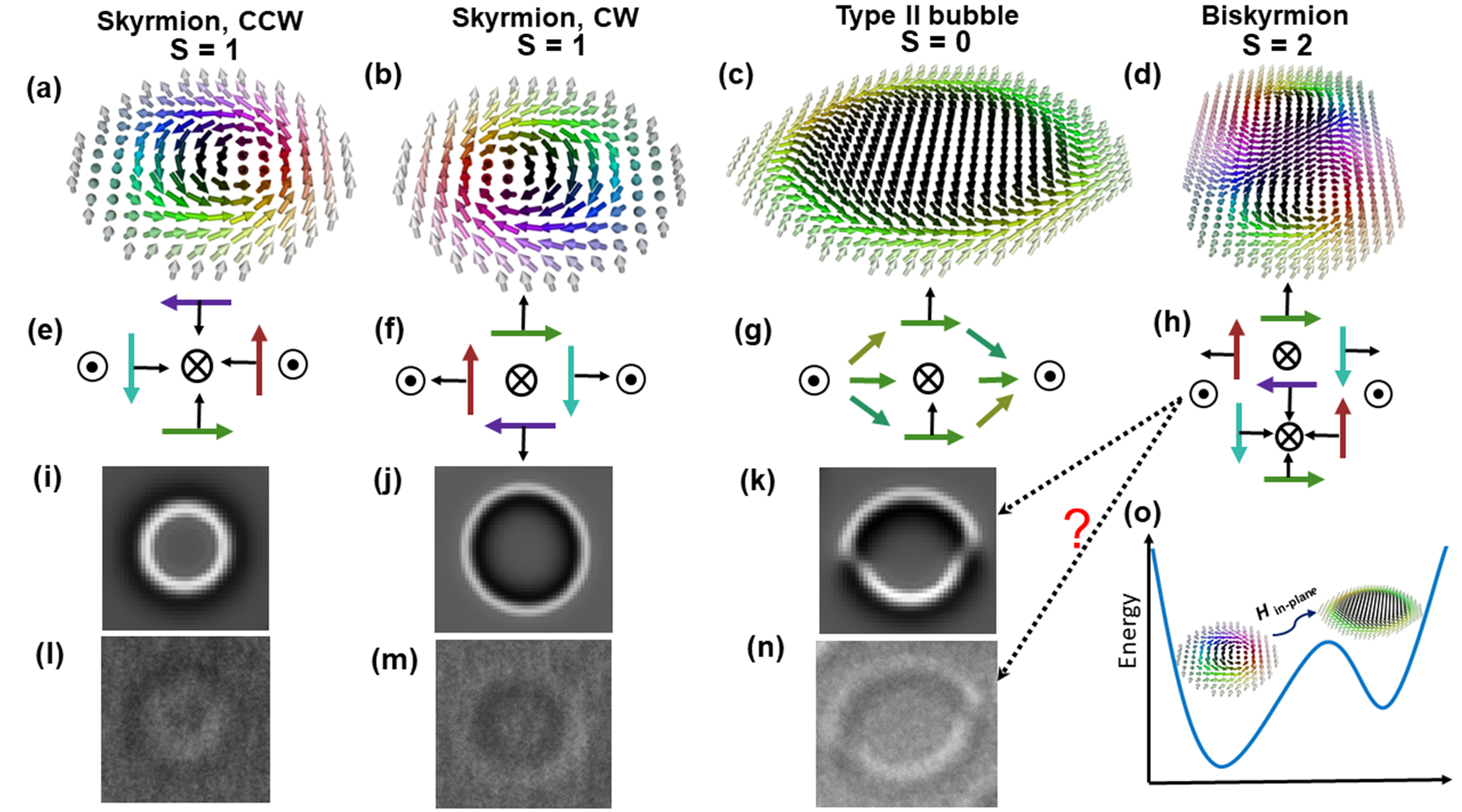}
		\caption{ Magnetic spin texture of (a)  skyrmion with counter-clockwise (CCW) helicity, (b)  skyrmion with clockwise (CW) helicity, (c) type-II bubble,  and (d) biskyrmion. (e)-(h) Schematics of the arrangement of in-plane magnetization components for the spin textures corresponding to (a)-(d), respectively. The black solid arrows in (e)-(h) represent the possible direction of electron deflection in LTEM. (i)-(k) The LTEM simulated images corresponding to the schematics in  (e)-(g). (l)-(n) Experimental LTEM images of the simulated LTEM pattern in (e)-(g). The dotted black arrows  from (h) to (k) and (n) point that the biskyrmion spin structure can also give rise to the same kind of LTEM contrast to that of type-II bubble. (O) Schematic diagram showing that both skyrmion and type-II bubble can exist in a material with skyrmion being the lowest energy state. Application of a small in-plane magnetic field can destabilize the skyrmion and  nucleate type-II bubble as the system lacks any particular chirality.}  
		\label{fig1}
	\end{center}
\end{figure*}

Along with the observation of skyrmionic bubbles, the finding of magnetic biskyrmions, which represent an addition of two skyrmions with opposite helicity, have also been reported in several centrosymmetric magnets\cite{LSMO0P315, MnNiGa, MnPdGa, NdCo5}. A summary of different spin textures reported  in the centrosymmetric magnets is depicted in Fig.~\ref{fig1}(a)-(n).  The 3D spin texture, as well as 2D schematics of the in-plane magnetization distribution of different possible magnetic bubbles are shown in Fig.~\ref{fig1}(a)-(h).  The Lorentz transmission electron microscopy (LTEM) simulated images of the spin textures are shown in Fig.~\ref{fig1}(i)-(k) and the corresponding experimental LTEM patterns are depicted in Fig.~\ref{fig1}(l)-(n). The skyrmions  with opposite helicity exhibit reverse black and white ring arrangement, whereas the type-II bubbles display alternative white and black half circles, as shown in Fig.~\ref{fig1}(k). It has  been shown that  the conventional type-II bubbles  [Fig.~\ref{fig1}(c) ] with topological charge zero can exhibit similar LTEM magnetic contrast like the  biskyrmions [Fig.~\ref{fig1}(d) ]  when viewed from a certain angle with respect to their axis \cite{type2bubble, type2bubble2}. Although there exists few recent reports highlighting the skyrmion to bubble transformation, we have carried out a systematic study on the aforementioned phenomenon  in a new skyrmion hosting hexagonal kagome ferromagnet Mn$_4$Ga$_2$Sn with skyrmion size of about 100~nm. We have mainly utilized the low temperature LTEM imaging technique to demonstrate a switching mechanism between the chiral skyrmions and the nonchiral type-II bubbles by systemetically tilting the sample in different directions to provide controlled in-plane magnetic field. In particular, we have shown that the stable magnetic skyrmions can be converted into the metastable type-II bubbles or vice versa by applying in-plane magnetic field, as schematically depicted in Fig.~\ref{fig1}(o).    

\section{RESULTS AND DISCUSSION}

\begin{figure*}
	\begin{center}
		\includegraphics[width=15cm]{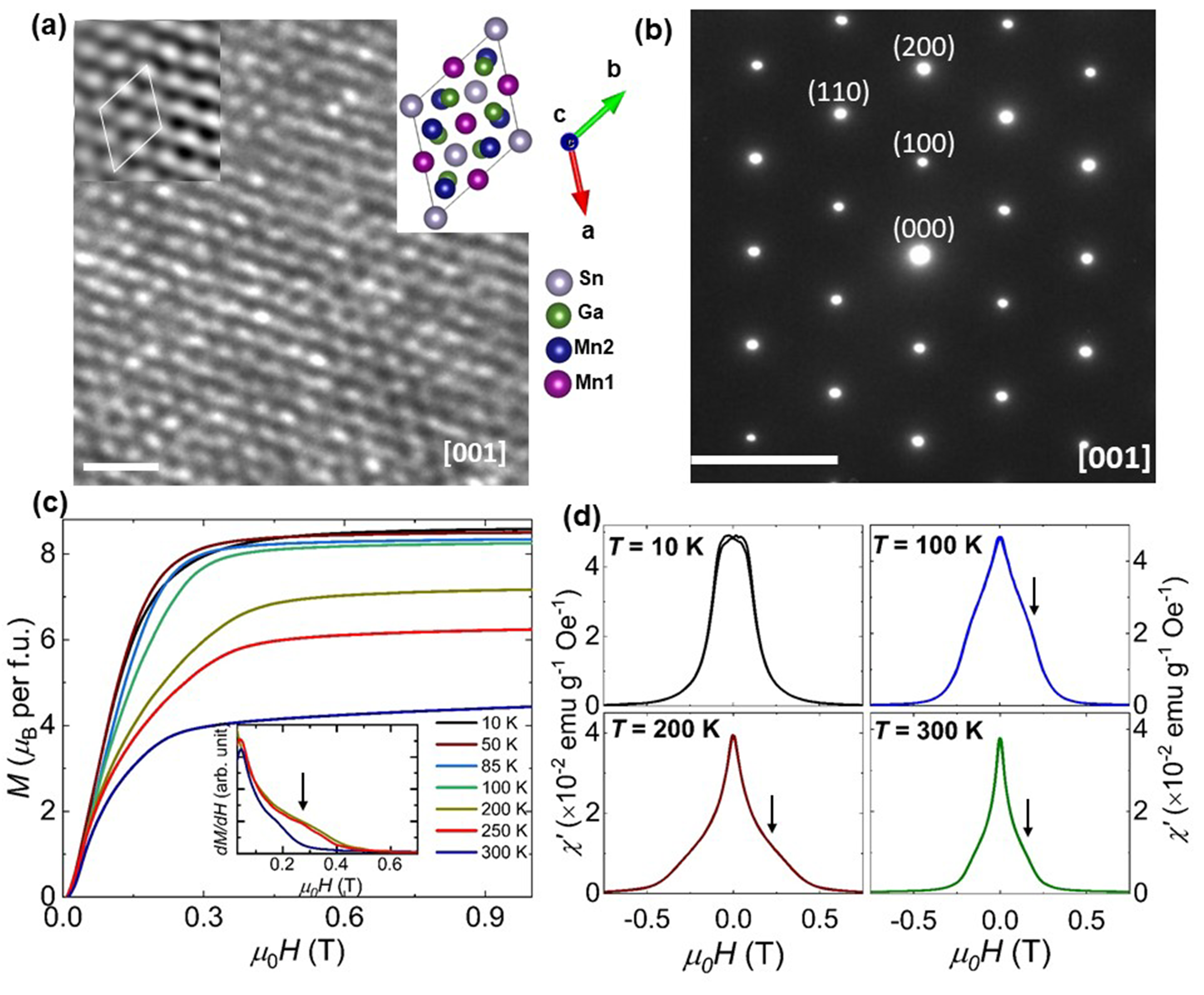}
		\caption{ (a) High resolution transmission electron microscopy (HRTEM) image taken on [001] oriented sample plate. The left inset shows the enlarged view of the atomic arrangement reconstructed by using the inverse fast Fourier transformation (iFFT) of the fast Fourier transformed HRTEM image. The right inset depicts the view of the atomic unit cell of Mn$_4$Ga$_2$Sn from the c-axis. The scale bar corresponds to 1 nm. (b) Room temperature selected area electron diffraction (SAED) pattern taken on the sample plate with [001] orientation.  The scale bar corresponds to 2 nm$^{-1}$. (c) Isothermal magnetization curves,  $M (H)$, measured at different temperatures. The inset shows the first derivative of $M (H)$ curves at selected temperatures. (d) ac susceptibility ($\textit{$\chi^\prime$})$ vs magnetic field ($\textit{$\mu_0$H}$) plots at different temperatures from 10~K to 300~K.  The black arrows point to the anomalies corresponding to the magnetic transitions. }
		\label{fig2}
	\end{center}
\end{figure*}


Mn$_4$Ga$_2$Sn crystallizes in Fe$_{6.5}$Ge$_{4}$-type hexagonal structure (space group P6$_3$/mmc) with alternative stacks of Mn-Sn and  Mn-Ga-Sn atomic layers arranged along the $c$-axis. A detailed analysis about the structural information is given in the Supplementary Fig.~\ref{S15}. We have performed high-resolution transmission electron microscopy (HRTEM) imaging on our thin plate sample as used for the LTEM study [Fig.~\ref{fig2}(a)]. The hexagonal unit cell formed by alternative Mn-Sn atoms is marked with a white box in the  inset of Fig.~\ref{fig2}(a) along with the corresponding crystal structure. Furthermore, the selected area electron diffraction (SAED) pattern shown in Fig.~\ref{fig2}(b) confirms the [001] orientation of our TEM lamella. The sample exhibits a Curie temperature ($T_{ C}$) of $\approx$ 320~K and an additional spin reorientation transition ($T_{ SR} $) at about 85~K ( see Supplementary Fig.~\ref{S18}). Field dependent isothermal magnetization, $M(H)$, measurements  are carried out at different temperatures to further access the magnetic state of our polycrystalline Mn$_4$Ga$_2$Sn sample [Fig.~\ref{fig2}(c)]. A large saturation magnetization of about 8.58 $ \mu_{\textrm B} $/f.u. is found at 10~K. A close look at the low field regime of the $M(H)$ curves measured at $ T > $ $T_{ SR} $ reveals the presence of kink kind of features that signify the existence of field induced magnetic phase transition in the system. This transition like characteristic can be clearly seen from the first-derivative of the $M(H)$ curves plotted in the inset of Fig.~\ref{fig2}(c). A similar type of transition anomaly has also been found in different skyrmion hosting materials  \cite{GaV4s8,Mnsi mh step,Gd2PdSi3}. For further verification of the observed transitions in the $M(H)$ measurements, we have carried out magnetic field dependent ac susceptibility measurements $\chi^\prime(H)$ at various temperatures from  $T$ = 10~K to $T$ = 300~K as depicted in Fig.~\ref{fig2}(d). The $\chi^\prime(H)$ data taken at 10~K exhibit a typical ferromagnetic like feature, whereas the measurement at 100~K shows an additional hump like anomaly that persists for the $\chi^\prime(H)$ data measured at 200~K and 300~K. It is important to mention here that the presence of transition like anomaly in the $\chi^\prime(H)$ data has been extensively used as a tool to indirectly probe the skyrmion phase in several skyrmion hosting materials \cite{Fe3Sn2,MnSi_acki,cu2OSeO3_acki,MnPtPdSn_sir,MnPtPdSn_jamal,Mn2NiGa}. Hence, the magnetic phase transitions found in the  isothermal magnetization as well as in the ac susceptibility measurements suggest the existence of a possible skyrmion phase above $T_{SR}$ in the present system.

\begin{figure*}
	\begin{center}
		\includegraphics[width=15cm]{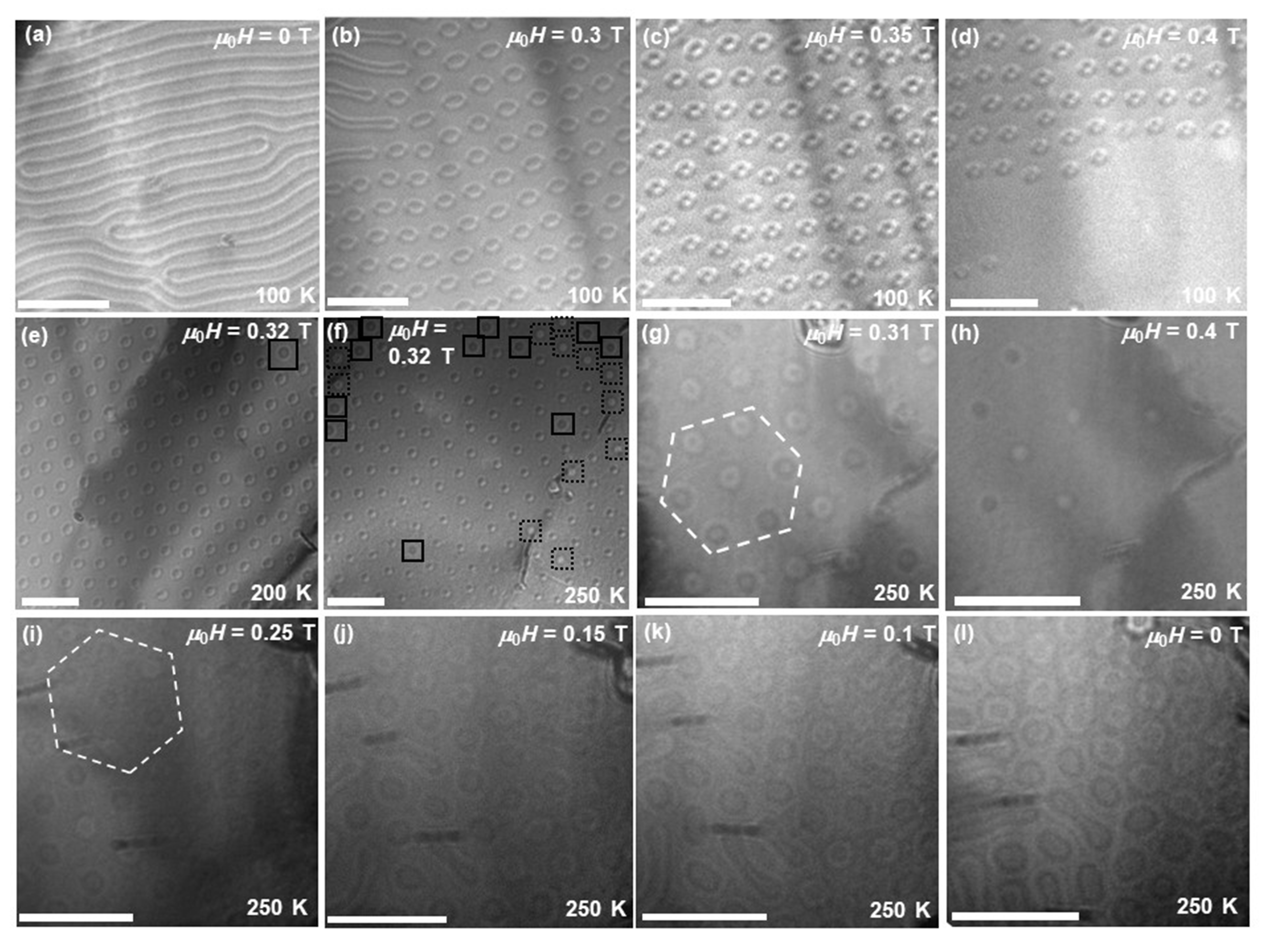}
		\caption{Over-focused LTEM images of  magnetic textures evolved with  magnetic fields and temperatures for Mn$_4$Ga$_2$Sn. (a-d) LTEM images of the magnetic states recorded  with increasing external magnetic field from 0~T to 0.4~T at 100~K. (e-f) LTEM images of the magnetic domain evolution at  magnetic field 0.32~T for temperatures 200~K and 250~K. The solid and dotted boxes represent skyrmionic textures with two different helicities. All the images in (a)-(f) are taken at nearly 4 degrees of sample tilting condition. (g-l) The field evolution of magnetic skyrmions at 250~K taken with increasing and decreasing the out-of-plane magnetic field at nearly zero tilting condition. The hexagonal skyrmion lattices are marked with the dotted hexagon in (g) and (i). The scale bars correspond to 500 nm.}
		\label{fig3}
	\end{center}
\end{figure*}

Motivated by the signature of field induced magnetic phase transitions in $M(H)$ and $\chi^\prime(H)$ data, we have performed an extensive real space LTEM imaging study at different temperatures. Figure~\ref{fig3}(a)-(d) shows over-focused LTEM images  recorded at 100~K by increasing the magnetic field from 0~T to 0.4~T. The presence of spontaneous stripe domains with an average period of about 100~nm are found at zero magnetic field [Fig.~\ref{fig3}(a)]. With increasing the fields, the stripe domains start to break into magnetic bubble like textures at a field of about 0.3~T [Fig.~\ref{fig3}(b)] and a lattice state is achieved at  $\approx$ 0.35~T [Fig.~\ref{fig3}(c)]. These magnetic textures  start to disappear around 0.4~T [Fig.~\ref{fig3}(d)] and  a field polarized state is obtained at $\mu_0H$ $\approx$ 0.5~T. The over-focused LTEM images taken at $T =$ 200~K and 250~K for $\mu_0H$ $\approx$ 0.32~T are depicted in  Fig.~\ref{fig3}(e)-(f). The field evolution of the magnetic states for these temperatures are shown in the Supplementary Fig.~\ref{S6}, \ref{S7},\ref{S8}. As it can be seen from Fig.~\ref{fig3}(e), a new type of magnetic contrast (shown inside the solid box) appears along with the magnetic textures found at 100~K.  By further increasing the temperature to 250~K, the emergence of additional magnetic contrast (shown in the dotted boxes) can be noticed together with the magnetic  states found at 200~K [Fig.~\ref{fig3}(f)]. Furthermore, the density of the newly formed magnetic textures greatly increases with increasing temperatures. A close look to the additional magnetic textures (shown inside the boxes) formed at 200~K and 250~K reveals the typical LTEM contrast of  skyrmions as previously shown in Fig.~\ref{fig1} (l) and (m). Similarly, the magnetic patterns found at 100~K corresponds to that of a type-II bubble [as shown in Fig.~\ref{fig1}(n)].

It is important to understand the stabilization mechanism that governs the formation of different magnetic states in our system. In order to avoid bending contours during the LTEM experiment in the present case, all the images shown in Fig.~\ref{fig3}(a)-(f) are taken by tilting the sample at a certain angle from the zone-axis ($\alpha$ = 1$^0$, $\beta$ = 3$^0$, where $\alpha$ and $\beta$ represent the sample tilting along $x$-axis and $y$-axis, respectively, which lie in the $ ab- $plane of the sample). As a result, a small in-plane magnetic field always acts along the basal plane of the sample in addition to the out-of-plane magnetic field. In this tilting condition, we find the existence of  type-II bubble  like magnetic contrast at 100~K. Interestingly, similar tilting condition at higher temperatures also gives rise to the finding of skyrmions with opposite helicity along with  type-II bubbles. The existence of strong bending contours at low temperatures up to 200~K restrict us to perform the LTEM imaging  along the zone axis. However, the  bending contour effect improves considerably at 250~K. To examine whether a non-zero in-plane magnetic field affects the nucleation of the observed magnetic textures,  we have recorded LTEM images at nearly zero tilting condition that makes sure the absence of any in-plane magnetic field in the sample. Fig.~\ref{fig3}(g)-(l) show the over-focused LTEM images  recorded by increasing and decreasing the out-of-plane magnetic fields at 250~K. Stripe domains with an average periodicity of 80~nm are observed as magnetic ground state (Supplementary Fig.~\ref{S9}). With increasing the magnetic fields, the stripe domains first start to shrink before skyrmion bubbles appear at a field of 0.15~T. Further increasing the field to 0.3~T helps in the formation of a hexagonal skyrmion lattice consisting of both CW and CCW spin rotation [marked with a hexagon in Fig.~\ref{fig3}(g)]. An applied field of 0.4~T leads to the survival of a few scattered skyrmions with a reduced size [Fig.~\ref{fig3}(h)]. Fig.~\ref{fig3}(i)-(l) show the evolution of skyrmion lattice with decreasing the magnetic fields starting from the saturated state. As it can be seen, a hexagonal skyrmion lattice can be nucleated upon decreasing the field to 0.25~T. Further decreasing the field to 0.1~T enables a mixed skyrmion and stripe domain textures, which remain as a remanent magnetic state at zero magnetic field. 

\begin{figure*}
	\begin{center}
		\includegraphics[width=15cm]{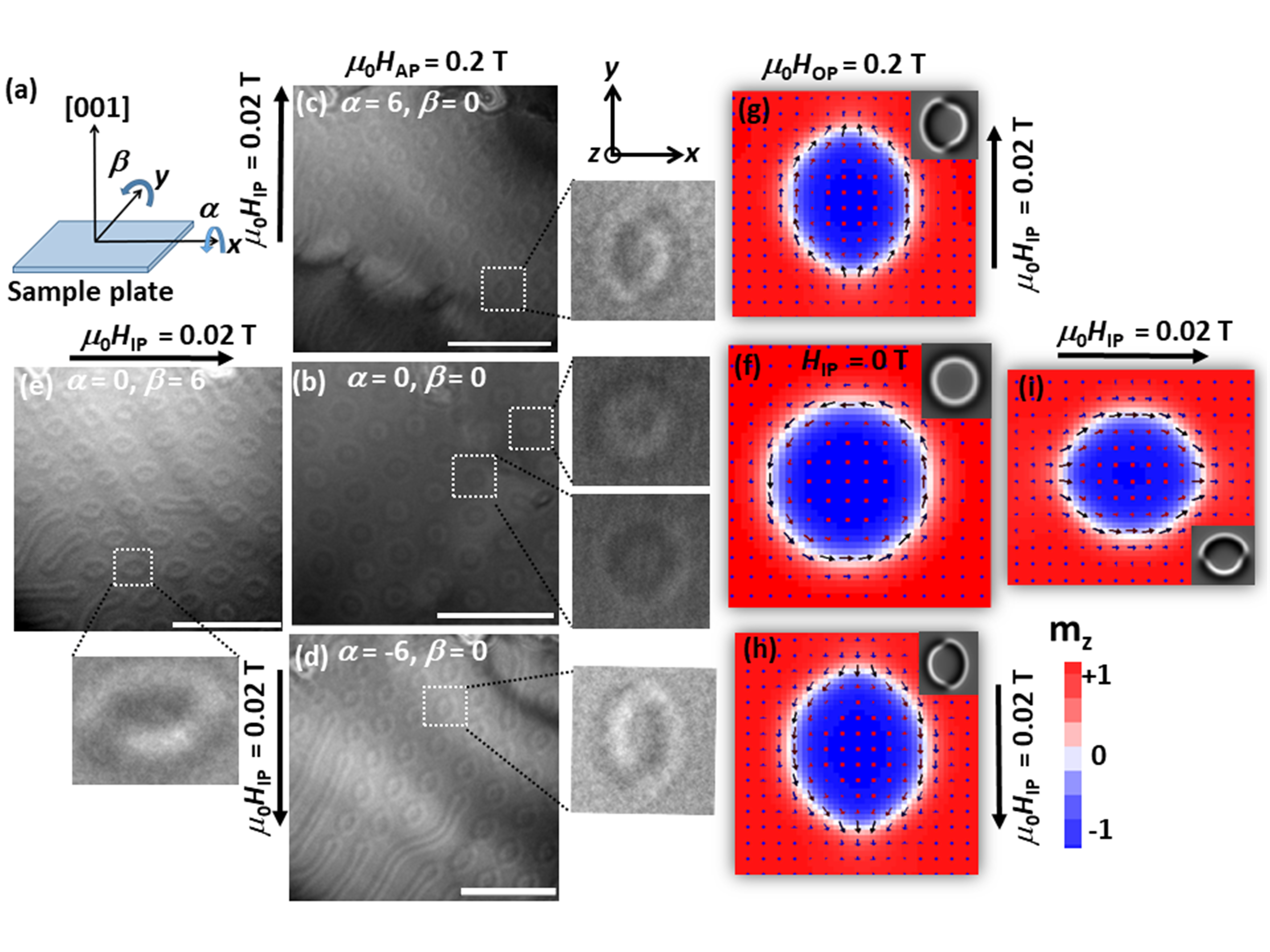}
		\caption{Evolution of magnetic domains with controlled tilting condition at a applied magnetic field (H$_{AP}$) of 0.2~T and temperature $T$ = 250~K. (a) Schematic diagram of the sample orientation to describe the tilting geometry. (b)-(e) The over-focused LTEM  images taken at different tilting angles. The scale bars correspond to 500 nm. (f)-(i) The OOMMF simulated magnetic structures with different in-plane magnetic field directions along with a fixed out-of-plane magnetic field of 0.2~T. The insets of (f)-(i) represent the LTEM simulated images corresponding to the OOMMF images. The extended figures corresponding to (f)-(i) are shown in the supplementary material.} 
		\label{fig4}
	\end{center}
\end{figure*}

Our LTEM imaging carried out with nearly zero tilting angle categorically demonstrates that the  skyrmions are the energetically favorable magnetic textures in the centrosymmetric magnets. The type-II magnetic bubbles are stabilized  when  a small  in-plane magnetic field is applied to the sample. For a better comprehension of the mechanism driving the nucleation of skyrmions/bubbles with in-plane magnetic fields, we have performed the LTEM imaging with controlled tilting conditions, as shown in Fig.~\ref{fig4}. The in-plane magnetic field is applied by tilting our sample around two orthogonal axes by angle $\alpha$ and $\beta$, as shown schematically in Fig.~\ref{fig4}(a). When the sample is tilted by $\alpha$ and $\beta$, it produces an in-plane magnetic field [$H_{IP}$] along the $y$ and $x$ directions, respectively. As it can be seen from Fig.~\ref{fig4}(b),  the over-focused LTEM image taken at $T$ = 250~K in zero tilting condition and  applied out-of-plane magnetic field of 0.2~T shows a hexagonal lattice of skyrmions  with both helicities. Introducing a 6$^{\circ}$ $\alpha$ tilting results in the formation of type-II bubbles as depicted in Fig.~\ref{fig4}(c). A close look at the individual bubble shown in the zoomed view of Fig.~\ref{fig4}(c) reveals that the deformation of the magnetic pattern exactly appears in the applied in-plane field direction ($ y-$axis).
Now, by reversing the tilting direction, i.e., - 6$^{\circ}$ $\alpha$ tilting,  the observed LTEM intensity contrast reverses its pattern as can be seen from Fig.~\ref{fig4}(d) and its zoomed view. This reversal of the LTEM intensity  represents a flipping of the in-plane magnetization direction of the underlying spin texture. Further introducing a rotation of the in-plane magnetic field, i.e., applying a 6$^{\circ}$ $\beta$ tilting, rotates the LTEM magnetic contrast by 90$^{\circ}$ as shown in Fig.~\ref{fig4}(e) and its zoomed view. These findings establish that the in-plane field is the deciding factor for the  internal magnetic structures of the observed magnetic textures.



To further support our experimental results, we have performed micromagnetic simulations using Object Oriented Micromagnetic Framework (OOMMF) by applying in-plane magnetic fields in different directions. The details of the simulation can be found in the Supplementary Fig.~\ref{S13}. Fig.~\ref{fig4}(f) shows one of the simulated spin textures obtained by applying 0.2~T out-of-plane magnetic field. The ring type pattern obtained from the LTEM simulation, shown in the inset of  Fig.~\ref{fig4}(f), corresponds to a Bloch-type skyrmion as observed experimentally. Now, we apply a small  in-plane magnetic field of 0.02~T along the $\pm y- $directions in addition to the out-of-plane magnetic field of 0.2~T.  The simulated spin textures resemble that of the type-II bubble as depicted inFig.~\ref{fig4}(g) and (h). It can be clearly seen that the observed spin textures exhibit an elongation along the applied in-plane field direction by breaking the clockwise/counter-clockwise spin rotation of the Bloch-type skyrmion. The LTEM simulated patterns shown in the insets of Fig.~\ref{fig4}(g) and (h) fully match with that of the experimental spin textures obtained for the $\alpha$ rotation given in Fig.~\ref{fig4}(c) and (d). When the in-plane magnetic field is changed to the $ x $-direction, the elongation of the spin texture follows the field direction [Fig.~\ref{fig4}(i)]. As expected, the simulated LTEM pattern shown in the inset of Fig.~\ref{fig4}(i) replicates the experimental spin texture observed in Fig.~\ref{fig4}(e). Hence, our experimental results along with the micromagnetic simulations, categorically establish that the magnetic textures obtained by application of small in-plane magnetic fields are indeed type-II bubbles.

\begin{figure}
	\begin{center}
		\includegraphics[width=9cm]{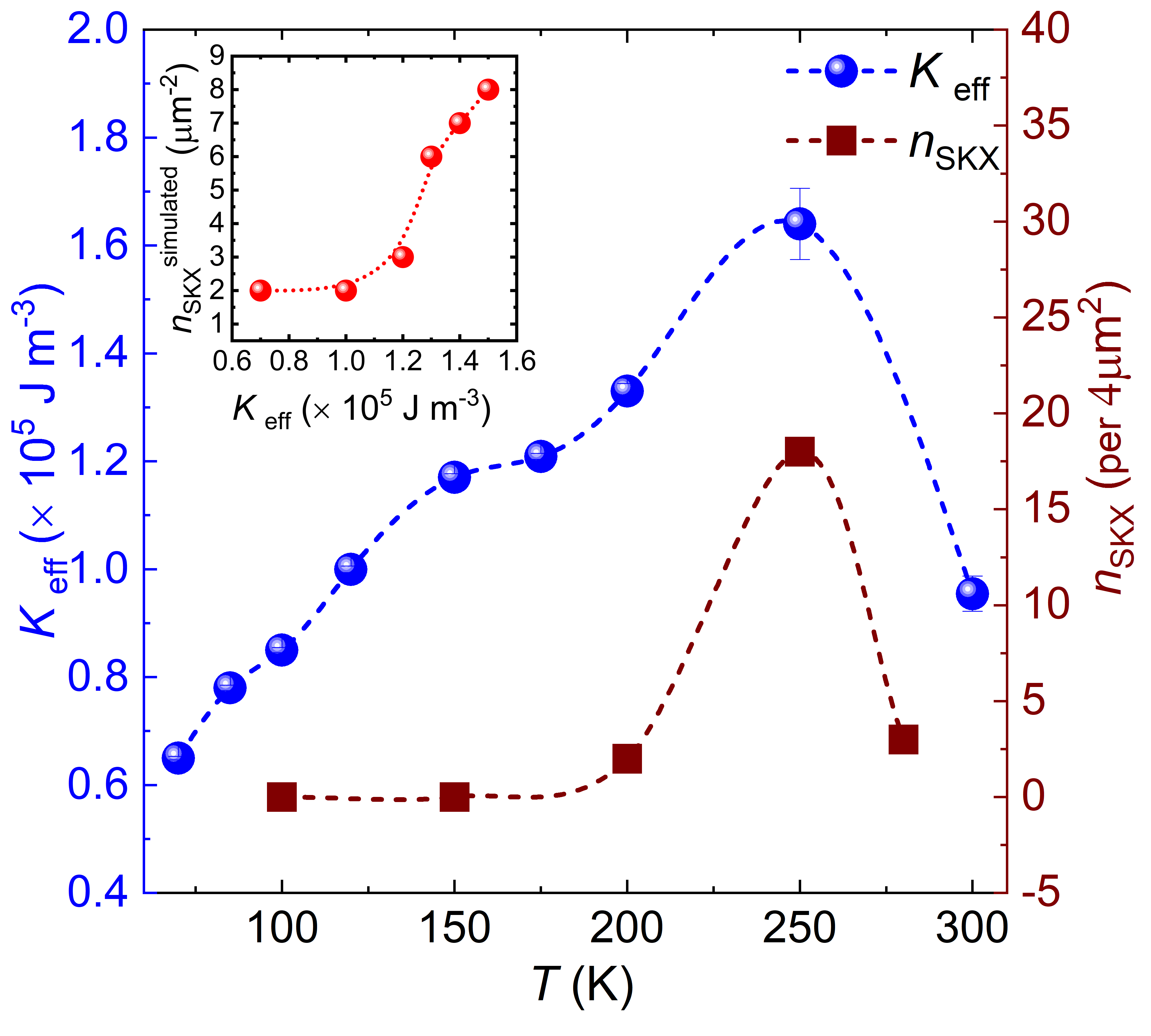}
		\caption{Temperature variation of effective magnetic anisotropy constant ($K_{eff}$) and number of skyrmions  ($n_{SKX}$) stabilized in tilted magnetic field in each 4~$\mu$m$^2$ area of the sample plate. The isolated skyrmion numbers up to 250~K are calculated for a field of 0.32~T, whereas at 280~K,  the  number is taken  for the field 0.2~T. The inset shows the number of simulated skyrmions in constant tilted magnetic field by varying the magnetic anisotropy (detailed simulation is given in the supplementary material).} 
		\label{fig5}
	\end{center}
\end{figure}

As it can be found from Fig.~\ref{fig3}, for a given in-plane magnetic field, the probability of skyrmion nucleation is higher at the higher temperatures, whereas type-II bubbles get easily nucleated at lower temperatures. This observation suggests that the effect of the in-plane magnetic field is more prominent at lower temperatures. It has been reported that Mn$_4$Ga$_2$Sn undergoes a transition from the high temperature easy-axis to  easy-plane  anisotropy at the $T_{ SR} $ \cite{Mn4Ga2Sn_thesis,Mn4Ga2Sn}. Since the magnetic anisotropy plays a major role in stabilizing skyrmions in centrosymmetric magnets, we have calculated the effective uniaxial anisotropy constant ($K_{eff}$) for Mn$_4$Ga$_2$Sn from $T$ = 70~K to $T$ = 300~K  based on the law of approach to saturation \cite{Anisotropy} (see Supplementary Fig.~\ref{S10} ). As it can be found from Fig.~\ref{fig5}, the effective  anisotropy exhibits its lowest value near the $T_{ SR} $ before increasing for the higher temperatures. The $K_{eff}$ attains a maxima at  $T$ $\approx$ 250~K, before decreasing rapidly at high temperatures as the $T_{ C}$ of the sample falls near the room temperature.

To examine the role of $K_{eff}$ on the stabilization of isolated skyrmions, the experimentally observed number of isolated skyrmions ($n_{SKX}$) are also plotted as a function of temperature in Fig.~\ref{fig5}. Interestingly, a direct correlation between the $n_{SKX}$ and $K_{eff}$  can be seen over the temperatures. To further substantiate our findings, we have carried out micromagnetic simulation to find out the evolution of skyrmions with different $K_{eff}$ in the presence of an in-plane magnetic field of 0.02~T (see the supplementary Fig.~\ref{S14}). The simulated number of isolated skyrmions ($n_{SKX}$$^{simulated}$)  as a function of $K_{eff}$  is shown in the inset of Fig.~\ref{fig5}. Corroborating our experimental results, the number of simulated isolated skyrmions increases with increasing $K_{eff}$. All these results suggest that stronger the $K_{eff}$ higher the probability of stabilizing the skyrmion phase. Hence, at low temperatures the in-plane magnetic field can easily nucleate the type-II bubble magnetic state by destabilizing the skyrmion phase at lower $K_{eff}$.

\begin{figure*}
	\begin{center}
		\includegraphics[width=16cm]{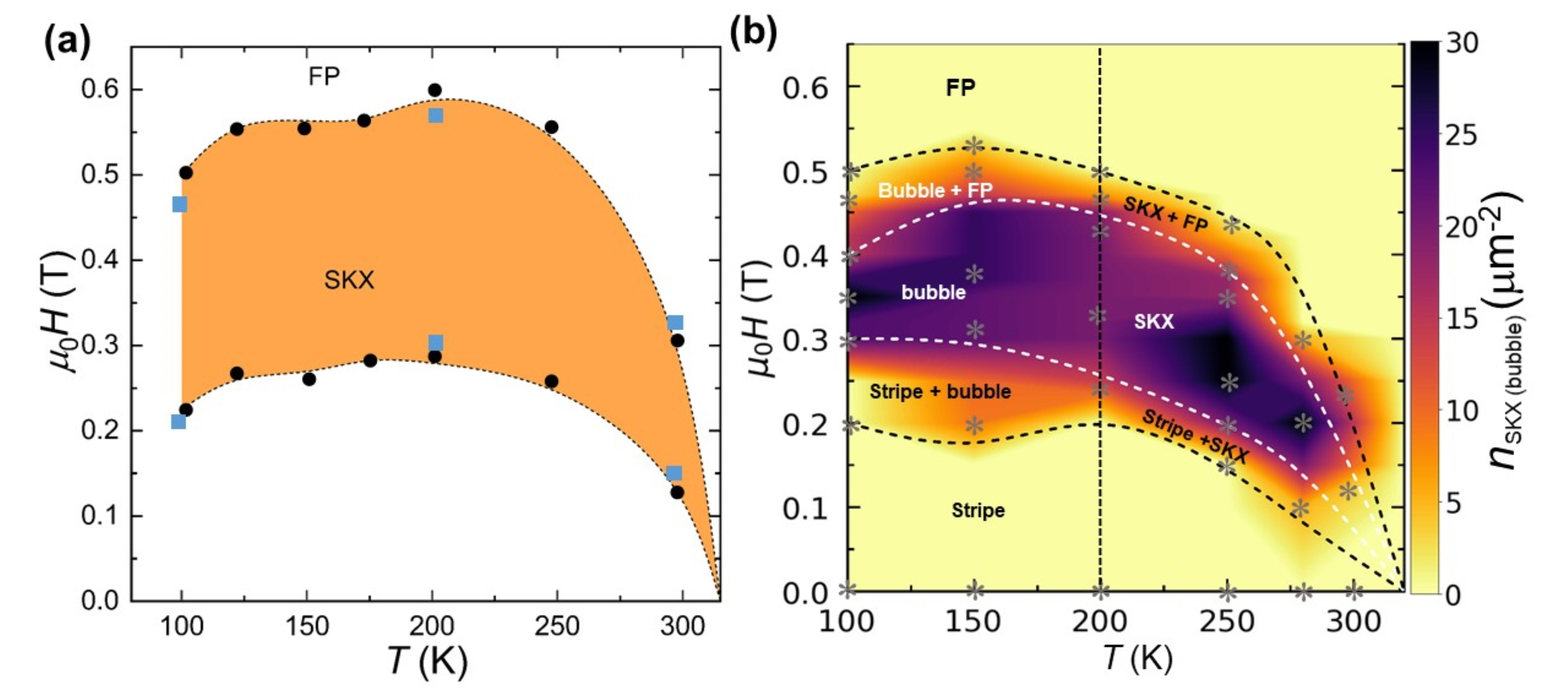}
		\caption{ (a) $H$ - $T$ phase diagram constructed using the magnetization data of polycrystalline Mn$_4$Ga$_2$Sn sample. The critical fields for constructing the phase diagram are taken from the point of anomaly in the $ M(H) $ measurements (filled circles) and ac susceptibility data (filled squares). (b) $ H-T $ phase diagram for Mn$_4$Ga$_2$Sn, as deduced from the temperature and field dependent LTEM measurements of the sample at nearly 4 degree tilting condition. The color contrasts represent the average numbers of skyrmions (SKX) and type-II bubbles in each $\mu$m$^2$ area. The asterisk symbols correspond to the point ($ H $ and $ T $) where images are recorded.} 
		\label{fig6}
	\end{center}
\end{figure*}



 Figure~\ref{fig6}(a) shows $H$-$T$ phase diagram drawn based on the ac suceptibility and dc magnetization measurements at different temperatures. The phase transition from ground state to the skyrmion phase is taken from the anomaly observed in the $ M(H) $ data obtained using the polycrystalline sample.
Based on our LTEM results, we have also constructed a phase diagram ($H$-$T$) as depicted in  Fig.~\ref{fig6}(b). We note that the present $H$-$T$ phase diagram is obtained for the case where we have to slightly tilt (nearly 4 degrees) the sample to get good quality images at all temperatures. With this sample tilting,  the type-II bubbles are the energetically favorable magnetic states at low temperatures. As discussed earlier, the transformation between the skyrmions and type-II bubbles can be achieved by  applying a non-zero in-plane magnetic field. Hence, it should be possible to realize the skyrmion phase instead of type-II bubbles at low temperatures. However, the bending contour effect at lower temperatures restricts us to perform LTEM experiment near the pole. The finding of nearly similar $H$-$T$ phase diagram for both the polycrystalline bulk sample and the single crystalline FIB lamella suggests that the observed anomaly in the magnetic signal of the polycrystalline sample indeed originates from the stripe domain to skyrmion phase transition.
 
Stabilization of different kinds of topological spin textures in the centrosymmetric materials have drawn a great attention because of their high degrees of freedom pertaining to use as multiple bits in data storage devices. 
Recently, nanomatric skyrmions of size about 1-2 nm size are observed in a few centrosymmetric systems, such as triangular lattice magnet Gd$_2$PdSi$_3$ \cite{Gd2PdSi3}, breathing kagome magnet Gd$_3$Ru$_4$Al$_{12}$ \cite{Gd3Ru4Al12} and tetragonal GdRu$_2$Si$_2$ \cite{GdRu2Si2}. The mechanism of skyrmion formation in these systems is attributed to the presence of  geometrical frustration and four-spin exchange interaction. In contrast, the skyrmions stabilized by competing dipolar and uniaxial magnetic anisotropy on other centrosymmetric systems are comparatively larger in size. However, skyrmions stabilized by geometrical frustration and higher order magnetic interactions are found to exists at very low temperatures in comparison to the room temperature skyrmions in other centrosymmetric skyrmion hosting materials. Even though there are few reports of room temperature skyrmions in the centrosymmetric magnets,  the size of the skyrmions  in these materials  happens to be greater than 200 nm \cite{LSFMO,Fe3Sn2}. In contrast, the present system hosts near room temperature  skyrmions  with size of about 100 nm. Besides the LTEM observations, the topological Hall effect (THE) measurements have also frequently been used as a signature for the skyrmion phase existance \cite{Gd2PdSi3,Mn2NiGa}. In general, the THE is inversely related to the size of the skyrmions. For the present  skyrmions with size  of 100 nm, the topological Hall resistivity should be in the order of nano-ohm-cm. The observation of THE in micro-ohm-cm order in the present system signifies that the THE mostly originates from the non-coplanar spin structure of the sample [see Supplementary Fig.~\ref{S19}, \ref{S20}].  

Recently, few reports show the transformation between topologically nontrivial skyrmions and trivial type-II bubbles in the centrosymmetric system \cite{Fe3Sn2_APL, Fe3Sn2_current control}. The present study makes a distinct effort to understand the underlying mechanism of skyrmion and bubble formation in centrosymmetric magnets. 
Similar kind of topological transition, e.g. antiskyrmion to non-topological bubble transformation, have also been experimentally observed in case of $ D_{2d} $ symmetry systems \cite{MnPtPdSn_tokura}, where the spin helix only stabilizes along the [100] or [010] direction. The anisotropic DM interaction (D$_X$ = - D$_Y$) in these systems dictates the helicity reversal of the antiskyrmions. Hence, the in-plane magnetic anisotropy in the $ D_{2d} $ skyrmion hosting system is only determined by the DM interaction.  In contrast, there is no preferential in-plane anisotropy in case of the centrosymmetric materials. The transformation between the topological skyrmions and the non-topological magnetic bubbles occurs due to nearly degenerate energies or small energy barrier between these spin textures. With the application of in-plane magnetic field, the Zeeman energy gain initiates the topological transformation easily. Hence, in the case of centrosymmetric system application of in-plane magnetic field in any direction gives rise to two Bloch lines with same in plane configuration. This results in the transformation of chiral skyrmion into achiral type-II bubble. When the in-plane magnetic field is removed the Bloch lines again disappear, enabling the transformation of achiral type II bubble to skyrmion. The reversal of spin textures with the applied in-plane magnetic field should be universal for all centrosymmetric skyrmion hosting materials where skyrmions stabilized by competing dipolar and uniaxial magnetic anisotropy. However, the magnitude of in-plane magnetic field may differ from system to system depending on the strength of the magnetic anisotropy in the system. Here, we for the first time show that the magnetic anisotropy of the system decides the magnitude of energy barrier for the transformation between the topological skyrmions and the non-topological magnetic bubbles. We systematically vary the in-plane magnetic field in different directions, both in our LETM measurements and micromagnetic simulations, to demonstrate the mechanism of skyrmion$ \leftrightarrow $bubble  transformation.  All our measurements categorically establish a direct correlation among the probability of skyrmion formation, in-plane magnetic field and magnetic anisotropy of the system. 


\section{CONCLUSION}

In conclusion, we have carried out an extensive LTEM measurement over a wide temperature range  to uncover the possible spin textures in the centrosymmetric kagome magnet Mn$_4$Ga$_2$Sn. Our experiment along with micromagnetic simulation conclusively establish that the Bloch-type skyrmions are the energetically most favorable  magnetic states in the hexagonal crystal based centrosymmetric magnets when the external magnetic field is applied along the magnetic easy-axis. In the case of LTEM imaging, it is extremely difficult to record magnetic contrast at the crystal zone axis due to the presence of bending contours. Hence, the application of an in-plane magnetic field  by a small tilting of the sample to avoid the bending contour can transform the skyrmionic state to the type-II bubble magnetic state.    We have also discussed the correlation between the skyrmion stabilization and change in the uniaxial anisotropy that plays an important role in skyrmion formation in these systems. Although skyrmions have been observed and well studied in many non-centrosymmetric systems, the skyrmion hosting centrosymmetric magnets having higher  degrees of freedom to manipulate their magnetic states deserve a thorough study for the realization of skyrmion based memory devices. 


\section{Method}
{\bf Sample preparation and structural characterization}: Polycrystalline ingots of Mn$_4$Ga$_2$Sn  were prepared using high pure Mn, Ga, Sn metals in the argon atmosphere by using an arc melting furnace. For further homogeneity, the ingots were sealed in a quartz tube in  the argon atmosphere and annealed at 823~K for ten days. The compositional homogeneity was verified using field emission scanning electron microscopy (FESEM) and energy-dispersive x-ray spectroscopy (EDX). The crystal structure of the sample was verified by  powder x-ray diffraction (XRD) measurements performed using Rigaku smartlab diffractometer with CuK$_{\alpha}$ radiation.

{\bf Magnetic measurements}: The DC magnetic measurements of the sample were performed by utilizing Superconducting Quantum Interference Device Vibrating Sample Magnetometer (SQUID-VSM, Quantum design). The ac-susceptibility measurements were carried out by using  the measurement ACMS option in Physical Properties measurement System (PPMS, Quantum design).  

{\bf Lorentz transmission electron microscopy (LTEM) study:} For the LTEM measurements, thin sample platelets of Mn$_4$Ga$_2$Sn with [001] orientation were cut from the polycrystalline sample using a Ga-based focused ion beam (FIB). The magnetic domains were studied using JEOL TEM (JEM-F200) in the Lorentz-TEM mode. A double tilting liquid nitrogen holder (GATAN-636) was used to study the temperature evolution of magnetic domains.  

{\bf Micromagnetic simulations}: Micromagnetic simulations were carried out using Object Oriented Micromagnetic Framework (OOMMF) code \cite{OOMMF}. Slab geometry of dimensions 1000 nm $\times$ 1000 nm $\times$ 100 nm were used for the simulation, with a rectangular mesh of 4 nm $\times$ 4 nm $\times$ 4 nm. The material parameters were chosen according to the experimental data of Mn$_4$Ga$_2$Sn. Exchange stiffness constant (A) was calculated using the formula, \textit{A} $\approx$ \textit{K$_B$T$_c$/a}, where \textit{K$_B$} is the Boltzman constant, \textit{T$_C$} is the Curie temperature and \textit{a} is the lattice constant of the sample. The estimated value of the exchange constant \textit{A} is about 5$\times$ 10$^{-12}$ J/m. The out-of-plane easy axis magnetocrystalline anisotropy (\textit{K$_u$}) at 250 K was calculated by using the formula \textit{D} = $\pi$ $\sqrt{\textit{A/K$_u$}}$, where D is the domain wall width observed in the LTEM experiment. The estimated value of magnetocrystalline anisotropy constant \textit{K$_u$} is about $\approx$ 1.02 $\times$ 10$^5$ J/m$^3$. The equilibrium states were obtained by fully relaxing the randomly distributed magnetization. The simulations for the magnetic field dependent domain evolution for 250 K and 100 K were conducted at zero temperature, with the experimental parameters are corresponding to 250 K to 100 K, respectively. We have used PYLorentz code to simulated the LTEM intensity contrasts corresponding to the OOMMF spin textures \cite{pylorentz}. To get the 3D spin textures, as shown in fig.~1, the spirit code software is used \cite{spirit code}.


AKN acknowledges the support from Department of Atomic Energy (DAE), the Department of Science and Technology (DST)-Ramanujan research grant (No. SB/S2/RJN-081/2016), SERB research grant (ECR/2017/000854) and Nanomission research grant
[SR/NM/NS-1036/2017(G)] of the Government of India.

\pagebreak

\onecolumngrid
\section{SUPPLEMENTARY INFORMATION}

\subsection{1.1. Sample crystal structure}
\setcounter{figure}{0}
\makeatletter 
\renewcommand{\thefigure}{S\@arabic\c@figure}
\makeatother
\begin{figure}[h]
	\includegraphics[angle=0,width=14cm,clip]{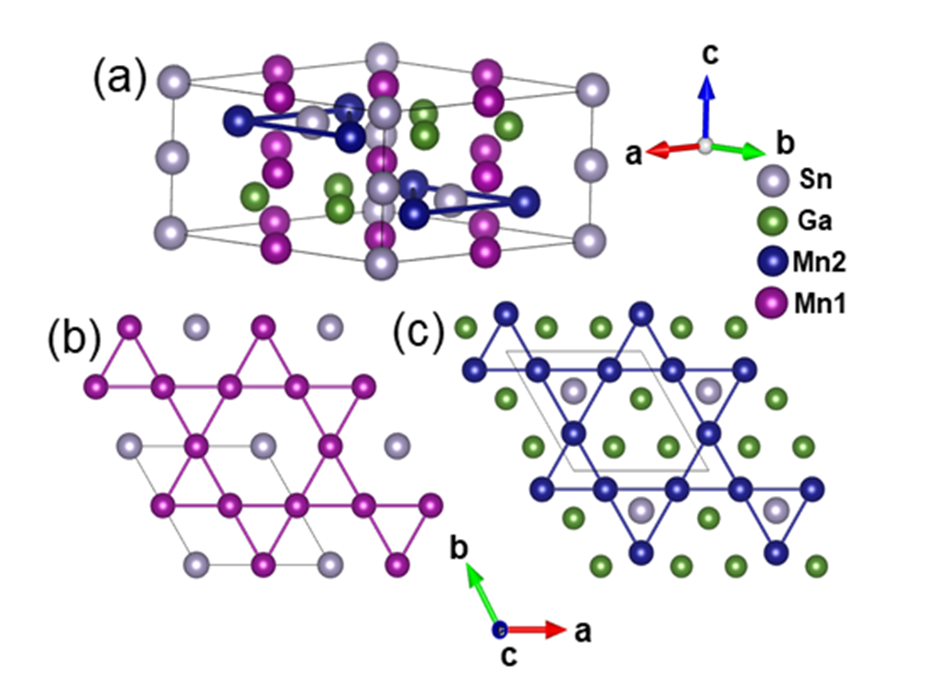}
	\caption{\label{S15} (a) The crystallographic unit cell of Mn4Ga2Sn. The top view (view from c-axis) of kagome lattice formed by Mn atoms in (b) Mn-Sn layer and (c) Mn-Ga-Sn layers. Grey, green, blue and pink balls represent the Sn, Ga, Mn2 and Mn1 atoms, respectively.}	
\end{figure}

Mn$_4$Ga$_2$Sn crystallizes in Fe$_{6.5}$Ge$_4$-type hexagonal structure (space group P63/mmc) with alternative stacks of Mn-Sn and Mn-Ga-Sn atomic layers arranged along the c-axis as depicted schematically in Fig.~\ref{S15}(a). The Mn atoms in the Mn-Sn layer form a kagome type of ordering with Sn sitting at the centre of the hexagon, whereas the Mn in the Mn-Ga-Sn layer exhibits a breathing kagome lattice where Sn atoms sit alternatively at the centre of the Mn triangles [ Fig.~\ref{S15} (b)-(c)].
\subsection{2. Sample characterization }

\subsection{2.1. X-ray diffraction (XRD)}

\begin{figure}[h]
	\includegraphics[angle=0,width=14cm,clip]{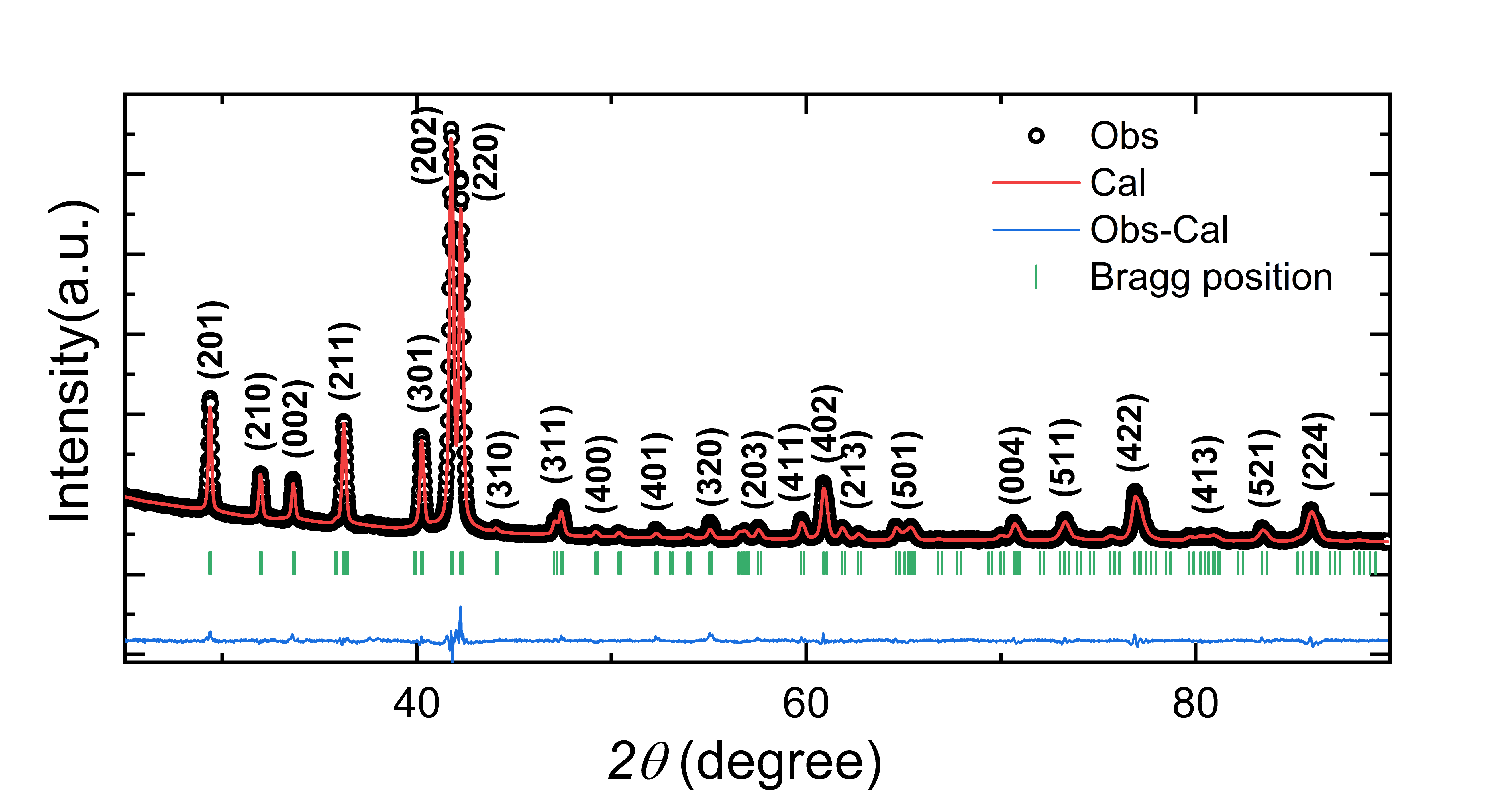}
	\caption{\label{S1} Room temperature powder XRD pattern for Mn$_4$Ga$_2$Sn sample. Black open circles represent the experimental data and the red solid line represent the simulated XRD pattern obtained from Rietveld refinement. The green scattered lines represent the Bragg position and difference between the experimental and simulated data is shown with the blue solid line.}	
\end{figure}

Room temperature powder XRD pattern with Rietveld refinement for Mn$_4$Ga$_2$Sn is shown in Figure~\ref{S1}.  The Rietveld refinement was performed using the space group P6$_3$/mmc.  In this structure the Mn atoms occupy two Wyckoff positions 6g (0.5, 0, 0) and 6h (x, 2x, 0.25) (x = 0.1562), whereas, Sn atoms preferentially occupy the 2a (0, 0, 0) and 2c (0.6667, 0.3333, 0.75) Wyckoff positions. The Ga atoms fully occupy the 6h (x, 2x, 0.75) (x = 0.1989) atomic position.  We have calculated the lattice parameters as a = b = 8.5527 \AA and c = 5.3264 \AA, which match well with the earlier reports \cite{Mn4Ga2Sn_thesis},\cite{Mn4Ga2Sn}.

\subsection{2.2. Scanning electron microscopy (SEM)}
\begin{figure}[h]
	\includegraphics[angle=0,width=10cm,clip]{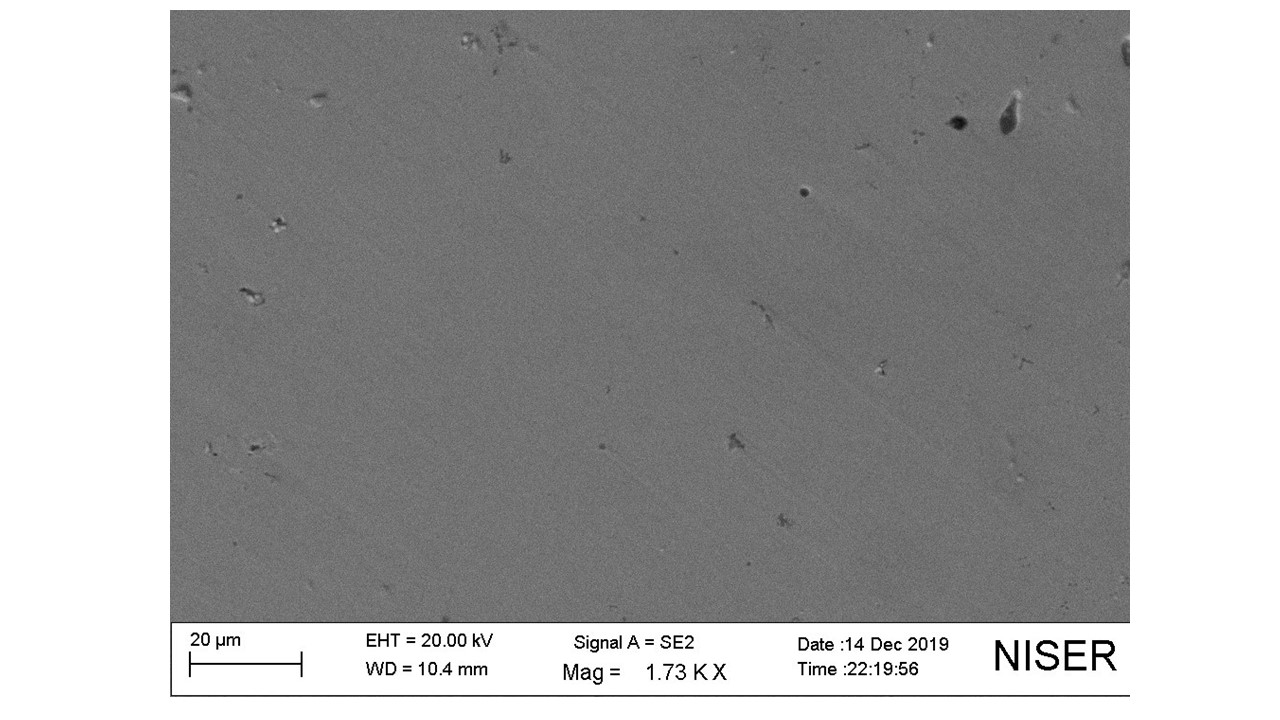}
	\caption{\label{S2} SEM image of Mn$_4$Ga$_2$Sn sample.}	
\end{figure}

The compositional homogeneity of the Mn$_4$Ga$_2$Sn sample is studied by using SEM and energy dispersive X-ray spectroscopy (EDX). The SEM image of the sample is shown in Figure~\ref{S2}.  The uniform contrast in the SEM image represents the single phase nature of our sample. The small black spots represent the small holes present in the sample. To determine the chemical composition of our sample we have performed EDX analysis at several places on our sample.  The observed EDX data nearly matches with the exact atomic composition of Mn$_4$Ga$_2$Sn as shown in Table~\ref{tab:table1}.

\begin{table}
	\begin{center}
		\caption{\label{tab:table1} Comparison between the exact atomic percentage and EDX obtained data.}
		\begin{tabular}{||c|c|c||}
		\hline
		Elements & Exact atomic percentage & obtained atomic percentage from EDX \\ [4.0ex]
		\hline\hline
		Mn & 57.14  &  58.76 $\pm$ 2.35 \\ [1.0ex]
		\hline
		Ga & 28.57  &  25.60 $\pm$ 1.27 \\ [1.0ex]
		\hline
		Sn & 14.29  &  14.27 $\pm$ 0.71 \\ [1.0ex]
		\hline
		\end{tabular}
	\end{center}
\end{table}

\subsection{3. Magnetic measurement}

\begin{figure}[h]
	\includegraphics[angle=0,width=12cm,clip]{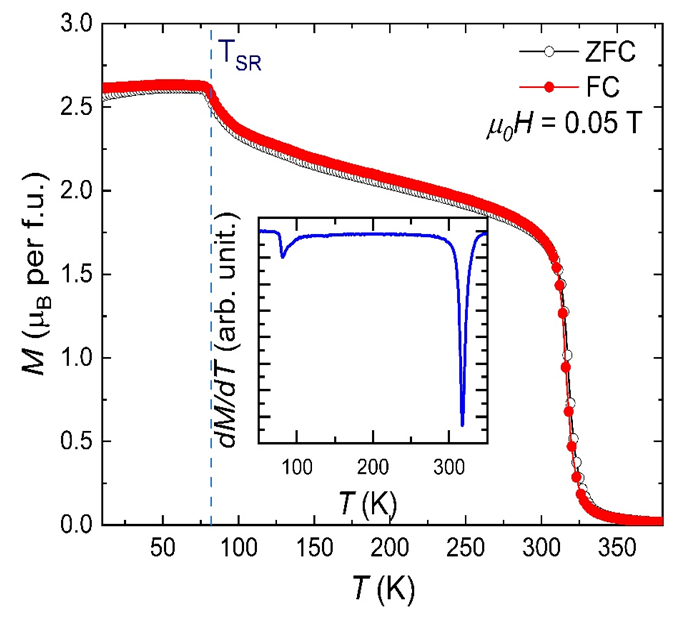}
	\caption{\label{S18} Temperature dependent magnetization $M(T)$ curves measured at $\mu_0$$H$ = 0.05 T in zero field cooled (ZFC) and field cooled (FC) modes. The inset shows the first derivative of the FC $M(T)$ curve [d$M(T)$/d$T$].}	
\end{figure}

In order to map out the magnetic state of the present Mn$_4$Ga$_2$Sn sample, we have conducted temperature dependent dc magnetization measurements [$M$($T$)] in zero field cooled (ZFC) and field cooled (FC) modes as shown in Fig.~\ref{S18}. The ZFC and FC $M(T)$ curves measured in a field of 0.05 T almost trace on each other with a small hysteretic behaviour for temperatures between 100~K and 300~K. As it can be found from the first derivative of the FC $M(T)$ curve with respect to temperature, shown in the inset of Fig.~\ref{S18}, the sample exhibits a Curie temperature ($T_C$) of 320~K and an additional transition at about 85~K.  This secondary transition at low temperature corresponds to a spin reorientation transition ($T_{SR}$), where the magnetic moments undergo a transition from easy-axis orientation at high temperatures to easy-plane alignment at the low temperature.  The effective anisotropy in the system continuously changes from negative (easy-plane) to positive value (easy-axis) with increasing temperatures. It is important to mention here that the mechanism of skyrmion formation in centrosymmetric magnets depends on the competing dipolar energy and easy axis magnetic anisotropy.

\subsection{4.	Lorentz transmission electron microscopy (LTEM)}

\begin{figure}[h]
	\includegraphics[angle=0,width=8cm,clip]{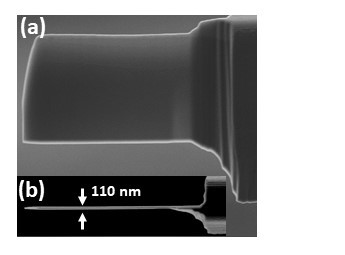}
	\caption{\label{S3} Images of FIB lamella of Mn$_4$Ga$_2$Sn sample. (a) Top view and (b) side view of sample lamella.}	
\end{figure}


\begin{figure}[h]
	\includegraphics[angle=0,width=15cm,clip]{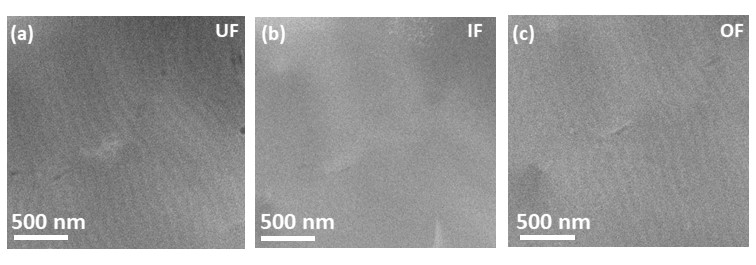}
	\caption{\label{S4} Room temperature LTEM images of Mn$_4$Ga$_2$Sn lamella in (a) under-focus (UF), (b) In-focus (IF) and (c) over-focus (OF) conditions at zero magnetic field.}	
\end{figure}


\begin{figure}[h]
	\includegraphics[angle=0,width=15cm,clip]{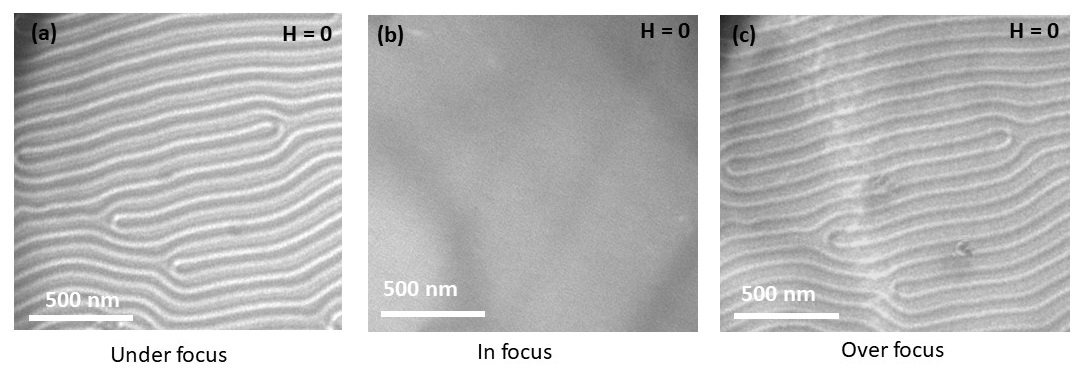}
	\caption{\label{S5} LTEM images of Mn$_4$Ga$_2$Sn lamella at 100 K in (a) UF, (b) IF and (c) OF modes at zero magnetic field.}	
\end{figure}


\begin{figure}[h]
	\includegraphics[angle=0,width=15cm,clip]{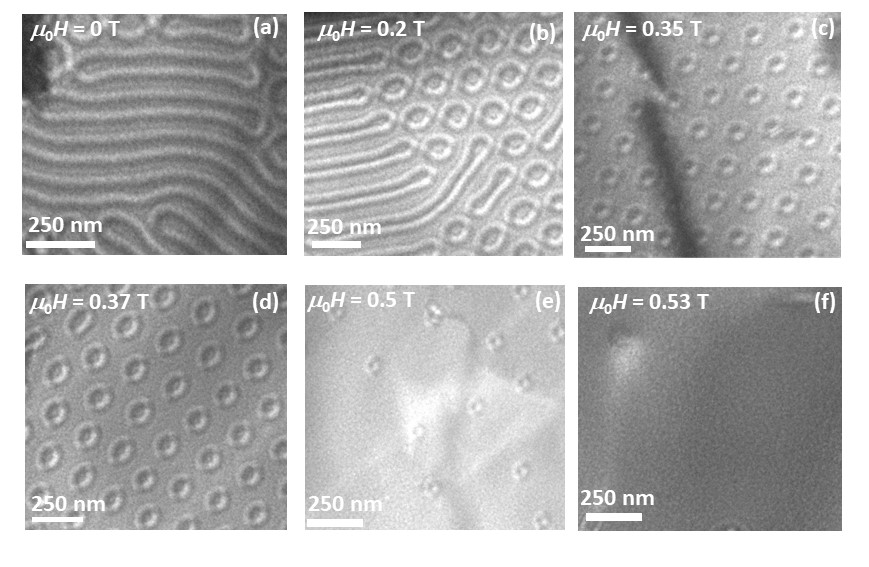}
	\caption{\label{S6} Over-focused LTEM images of Mn$_4$Ga$_2$Sn sample lamella (with 4 degree sample tilting) at 150 K at different magnetic fields from $\mu_0$$H$ = 0 T to $\mu_0$$H$ = 0.53 T.}	
\end{figure}

\begin{figure}[h]
	\includegraphics[angle=0,width=15cm,clip]{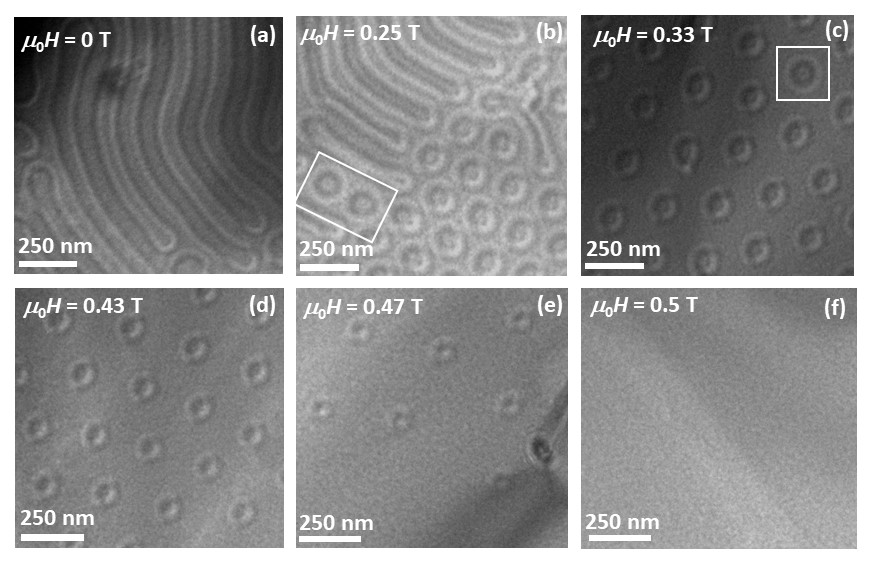}
	\caption{\label{S7} Over-focused LTEM images of Mn$_4$Ga$_2$Sn sample lamella (with 4 degree sample tilting) at 200 K at different magnetic fields from $\mu_0$$H$ = 0 T to $\mu_0$$H$ = 0.5 T.}	
\end{figure}

\begin{figure}[h]
	\includegraphics[angle=0,width=15cm,clip]{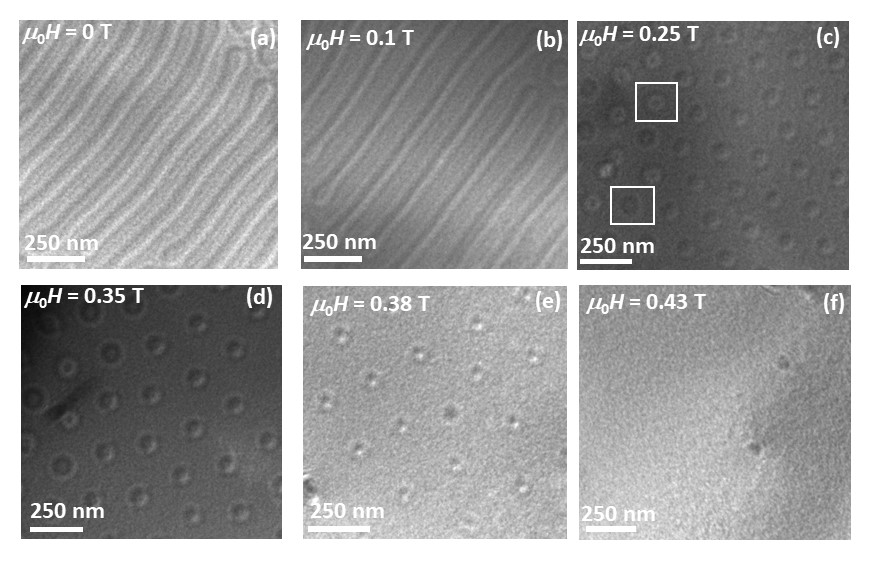}
	\caption{\label{S8} Over-focused LTEM images of Mn$_4$Ga$_2$Sn sample plate (with 4 degree sample tilting) at 250 K at different magnetic fields from  $\mu_0$$H$ = 0 T to  $\mu_0$$H$ = 0.43 T. }	
\end{figure}


\begin{figure}[h]
	\includegraphics[angle=0,width=15cm,clip]{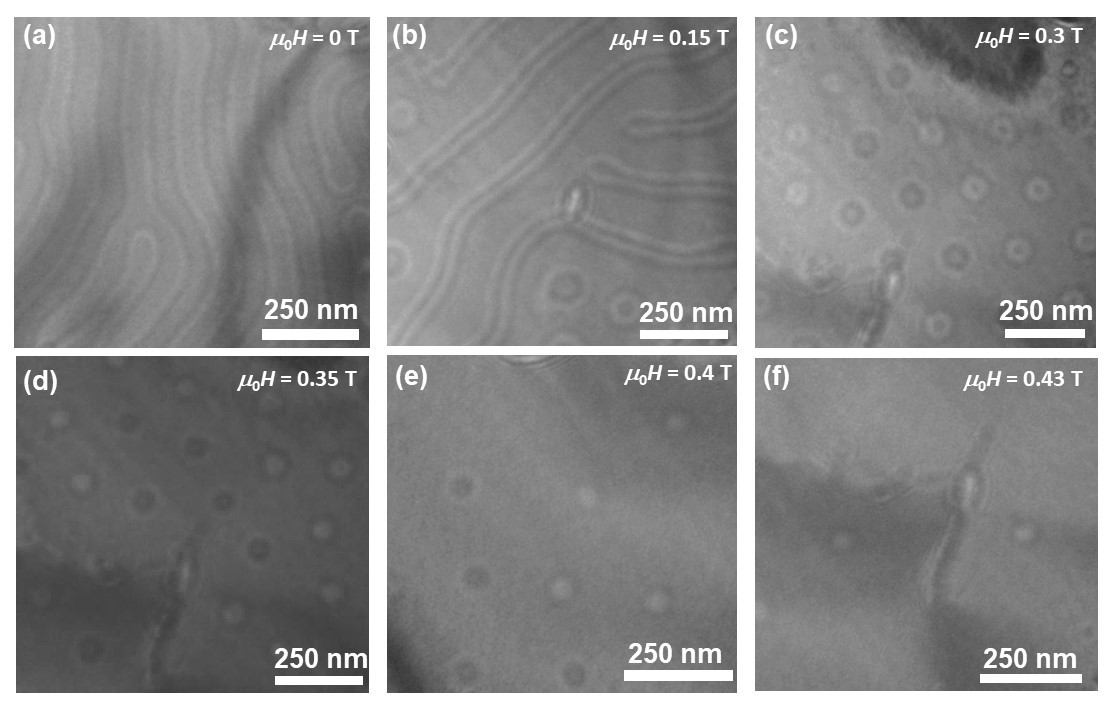}
	\caption{\label{S9} Over-focused LTEM images of Mn$_4$Ga$_2$Sn sample plate (with nearly 0 degree sample tilting) at 250 K at different magnetic fields from  $\mu_0$$H$ = 0 T to $\mu_0$$H$ = 0.43 T.}	
\end{figure}

For the LTEM measurements we have prepared 110 nm thin plate of Mn$_4$Ga$_2$Sn sample using Ga-based focused ion beam (FIB) technique.  Figure~\ref{S3} shows the FIB image of the sample lamella from top view and side view.

As our sample exhibits a Curie temperature of about 320 K, very poor LTEM magnetic contrast is found at room temperature as shown in Figure~\ref{S4}. To understand the magnetic texture with better contrast we have cooled down our sample in cryogenic temperature ($T$ = 100 K) using liquid-N2 TEM sample holder. Figure~\ref{S5} shows the magnetic ground state of the sample at 100 K.

Magnetic field dependent LTEM imaging were also performed at various temperatures from $T$ = 100 K to $T$ = 300 K. Variation of magnetic domains with change in magnetic field at temperatures of 150 K, 200 K and 250 K are shown in Figure~\ref{S6}, \ref{S7} and \ref{S8}, respectively. Here, along with the type-II bubbles, we have observed a larger number of isolated skyrmions at a temperature about 250 K, as marked with white boxes. The field sweep measurement at 250 K with zero tilting condition is also shown in Figure~\ref{S9}. At zero tilting condition formation of skyrmionic bubble from stripe domain is observed.


\subsection{5.	Micromagnetic simulation}

\begin{figure}[h]
	\includegraphics[angle=0,width=16cm,clip]{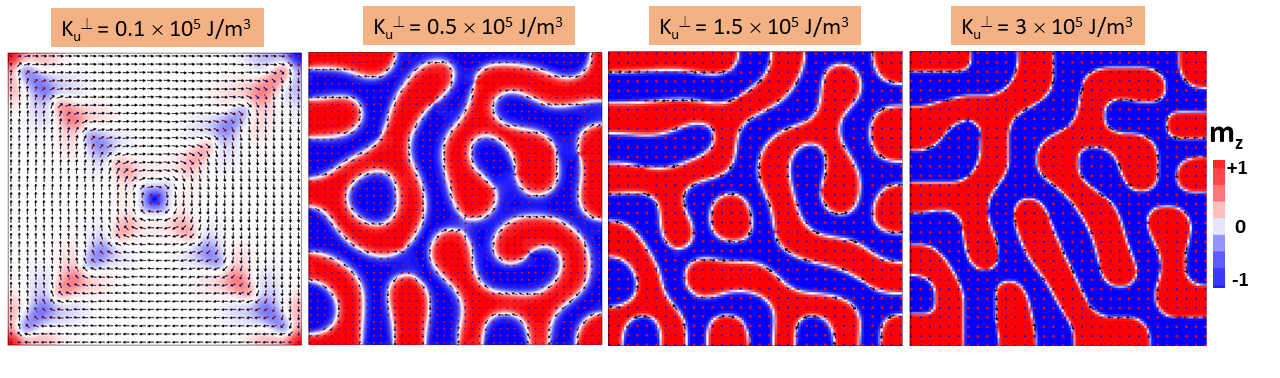}
	\caption{\label{S11} The simulated magnetic ground states for different uniaxial anisotropy constants, with the sample parameters corresponding to $T$=250~K, $A$ = 5 $\times$ 10$^{-12}$~J/m, $M_s$ = 5.35 $\times$ 10$^5$~A/m. The simulation is performed considering sample area 1$\mu$m $\times$ 1$\mu$m and thickness 100~nm.}	
\end{figure}

\begin{figure}[h]
	\includegraphics[angle=0,width=16cm,clip]{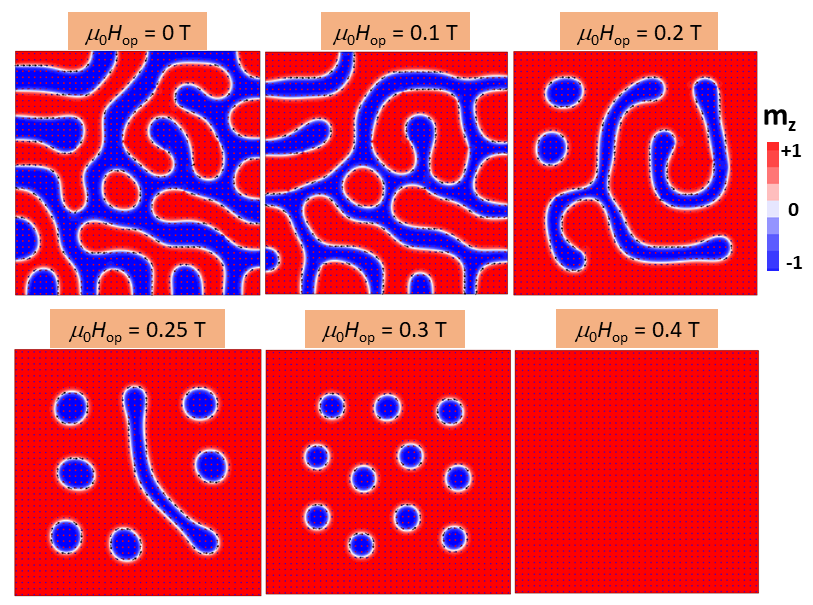}
	\caption{\label{S12} The simulated magnetic structures at different out-of-plane magnetic fields, with the sample parameters corresponding to 250~K. The used parameters are, $A$ = 5 $\times$ 10$^{-12}$ J/m, $M_s$ = 5.35 $\times$ 10$^5$ A/m, $K_u^\perp$= 1.0 $\times$ 10$^5$ J/m$^3$. }	
\end{figure}

\begin{figure}[h]
	\includegraphics[angle=0,width=16cm,clip]{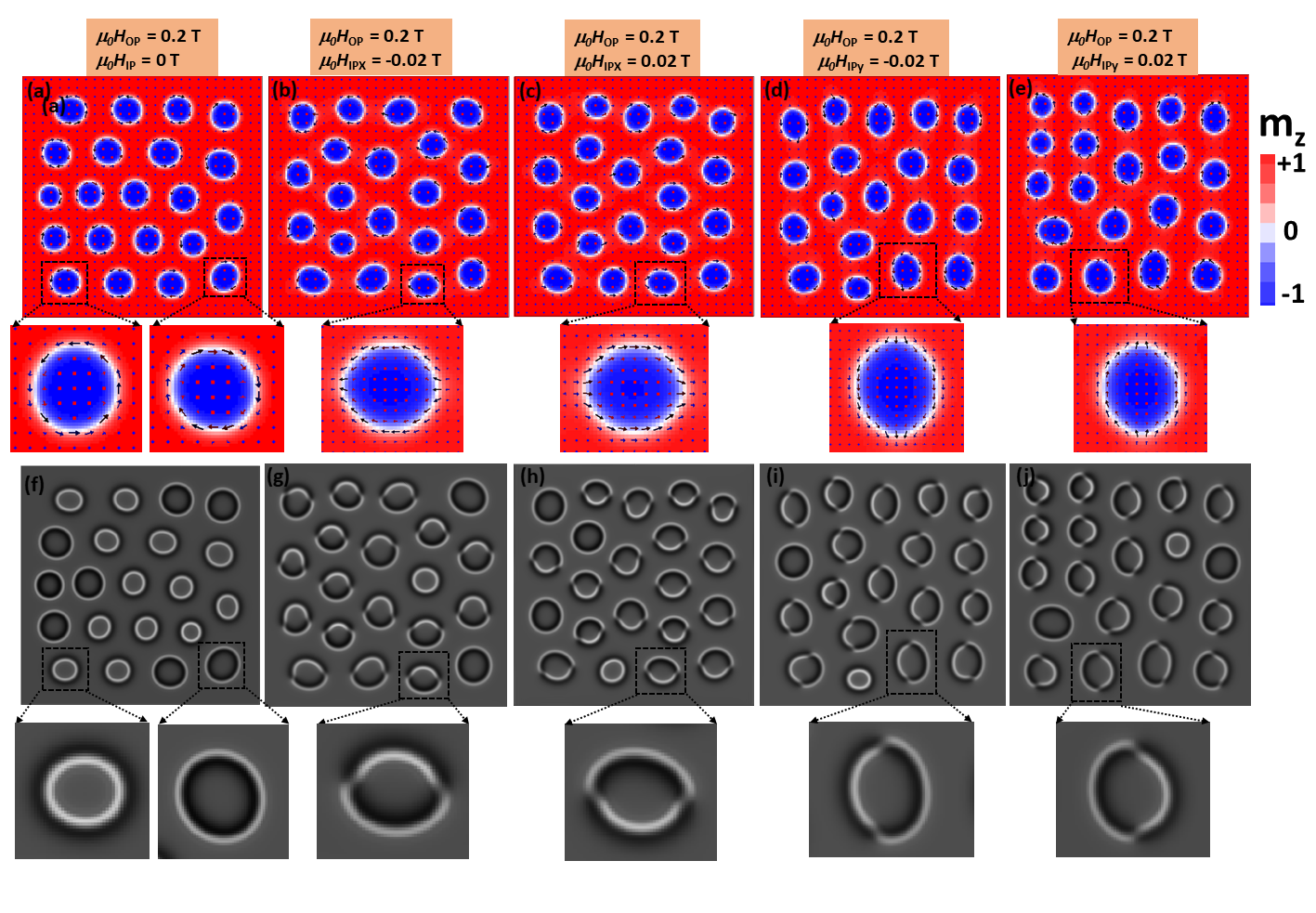}
	\caption{\label{S13} The simulated magnetic structures with out-of-plane magnetic field ($\mu_0$$H_{OP}$) of 0.2~T and in-plane magnetic field ($\mu_0$$H_{IP}$) of 0.02~T at different directions (a)-(e).  The simulations are performed with the sample parameters corresponding to 250~K. The used parameters are, $A$ = 5 $\times$ 10$^{-12}$ J/m, $M_s$ = 5.35 $\times$ 10$^5$ A/m, $K_u^\perp$= 1.0 $\times$ 10$^5$ J/m$^3$.. All the equilibrium states are achieved after fully relaxing the random magnetic state at that particular conditions. The simulation is performed considering sample area 1$\mu$m $\times$ 1$\mu$m and thickness 100~nm. The LTEM simulations of the magnetic structure corresponding to the figure (a) – (e) in over-focused mode is performed using PYLorentz code (f) – (j). }	
\end{figure}

We have carried out a detailed micromagnetic simulation to understand the mechanism of type-II bubble and skyrmion formation in our sample.  The micromagnetic simulations is carried out using the Object Oriented Micromagnetic Framework (OOMMF) programme\cite{OOMMF}. The equilibrium states are obtained by fully relaxing the random magnetic state. It is found that the uniaxial magnetocrystalline anisotropy has a fundamental role for the stabilization of stripe domain as magnetic ground state. For very low anisotropy the vortex ground state is observed rather than stripe domain state.

\begin{figure}[h]
	\includegraphics[angle=0,width=16cm,clip]{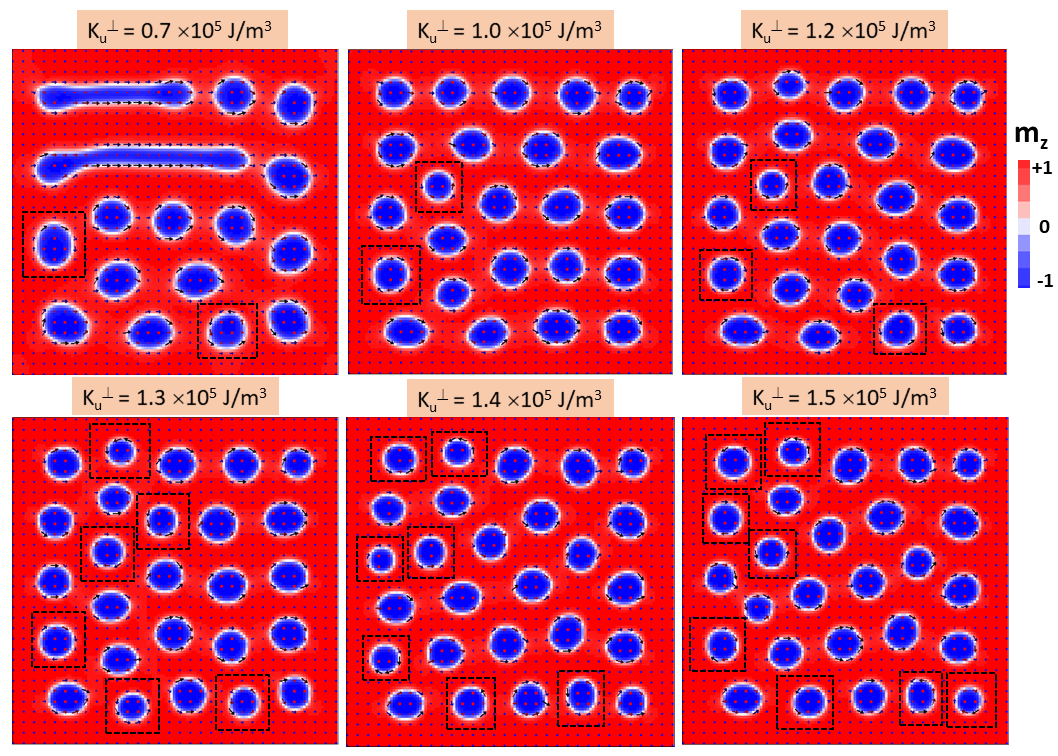}
	\caption{\label{S14} The simulated magnetic textures for different uniaxial anisotropy constant values with out of plane magnetic field of 0.3~T and in-plane magnetic field of 0.02~T. The used parameters are, $A$ = 5 $\times$ 10$^{-12}$ J/m,$M_s$ = 6.68$\times$ 10$^5$ A/m (saturation magnetization corresponding to 100~K). The black arrows represent the in-plane magnetization components. The simulation performed considering sample area 1$\mu$m $\times$ 1$\mu$m and thickness 100~nm. The skyrmion spin textures are marked with black dashed boxes. }	
\end{figure}

\begin{figure}[h]
	\includegraphics[angle=0,width=16cm,clip]{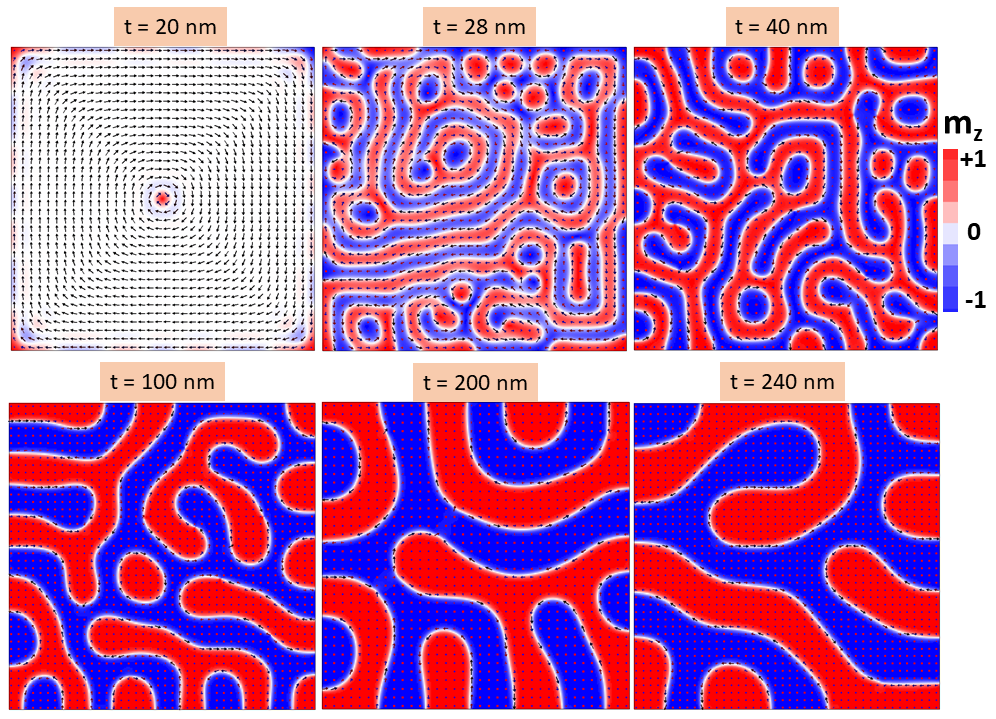}
	\caption{\label{S17} The simulated magnetic ground state for different sample thickness, used parameters are, A = 5 $\times$10-12 J m$^{-1}$, $M_s$ = 5.35 $\times$10$^5$ A m$^{-1}$ (saturation magnetization corresponding to 250~K), $K_u^\perp$= 1.0 $\times$10$^5$ J m$^{-3}$. The black arrows represent the in-plane magnetization components. Colour bar represents the magnetic moment along $z$-direction. The simulation is performed considering sample area 1$\mu$m $\times$1$\mu$m. All the equilibrium states are achieved after fully relaxing the random magnetic state at that particular conditions.  }	
\end{figure}

\begin{figure}[h]
	\includegraphics[angle=0,width=16cm,clip]{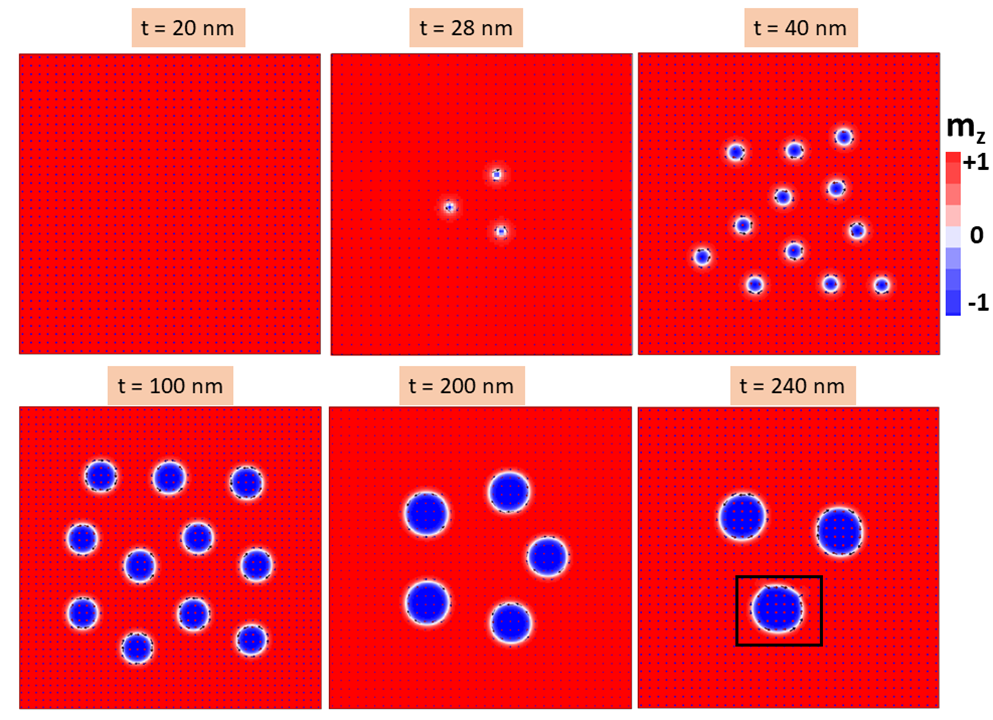}
	\caption{\label{S16} The simulated magnetic state at an out of plane magnetic field of 0.3 T for different sample thickness, used parameters are, A = 5 $\times$10-12 J m$^{-1}$, $M_s$ = 5.35 $\times$10$^5$ A m$^{-1}$ (saturation magnetization corresponding to 250~K), $K_u^\perp$= 1.0 $\times$10$^5$ J m$^{-3}$. The black arrows represent the in-plane magnetization components. Colour bar represents the magnetic moment along $z$-direction. The simulation is performed considering sample area 1$\mu$m $\times$1$\mu$. All the equilibrium states are achieved after fully relaxing the ground magnetic state at that particular conditions. The type-II bubble spin textures are marked with black box.  }	
\end{figure}

The magnetic field dependent micromagnetic simulations is carried out using the sample parameters corresponding to 250 K and considering the stripe domain as initial state at zero magnetic field (Figure~\ref{S11}). As can be seen in Figure~\ref{S12}, the stripe domains are transformed into magnetic skyrmions with both clockwise and counter-clockwise helicities with increase in magnetic field.

The micromagnetic simulations are also performed for various sample tilting conditions i.e, with application of additional in-plane magnetic field in different directions. We have observed almost all the skyrmions are transformed to type II bubble with application of additional in-plane magnetic field corresponding to the 6-degree sample tilting condition with respect to x-axis and y-axis, as shown in Figure~\ref{S13}.

To understand the effect of uniaxial magnetocrystalline anisotropy on the skyrmion to type-II bubble transformation with in-plane magnetic field, the spin textures are simulated with varying uniaxial anisotropy ($K_u^\perp$) as shown in Figure~\ref{S14}.  The simulations are carried out with an out-of-plane magnetic field of 0.3 T and an in-plane magnetic field of 0.02 T. The number of isolated skyrmions per $\mu$m$^2$ area increases with increase in the uniaxial magnetocrystalline anisotropy. 

To understand the effect of dipolar energy on the underlying spin textures we have simulated the magnetic states with varying sample thickness, as dipolar energy can be varied by changing the sample thickness. We have observed that with decreasing sample thickness i.e., increasing dipolar energy, the magnetic ground state changes from the out-of-plane stripe domains to in-plane vortex domain [Fig.~\ref{S17}]. As the dipolar energy always prefers in plane spin arrangement, with dominant dipolar energy contribution at low sample thickness enables the vortex domain state, whereas increasing the thickness stabilizes stripe domains with larger width. With application of a magnetic field of 0.3 T transforms the stripe domains to skyrmion lattice for sample thickness up to 200 nm [Fig.~\ref{S16}]. For the sample thickness greater than 240 nm, we find few type-II bubble along with skyrmion. Our simulation also shows that increase in sample thickness enhances the size of the skyrmions. 
\subsection{5. Calculation for temperature dependent effective magnetocrystalline anisotropy}

\begin{figure}[h]
	\includegraphics[angle=0,width=16cm,clip]{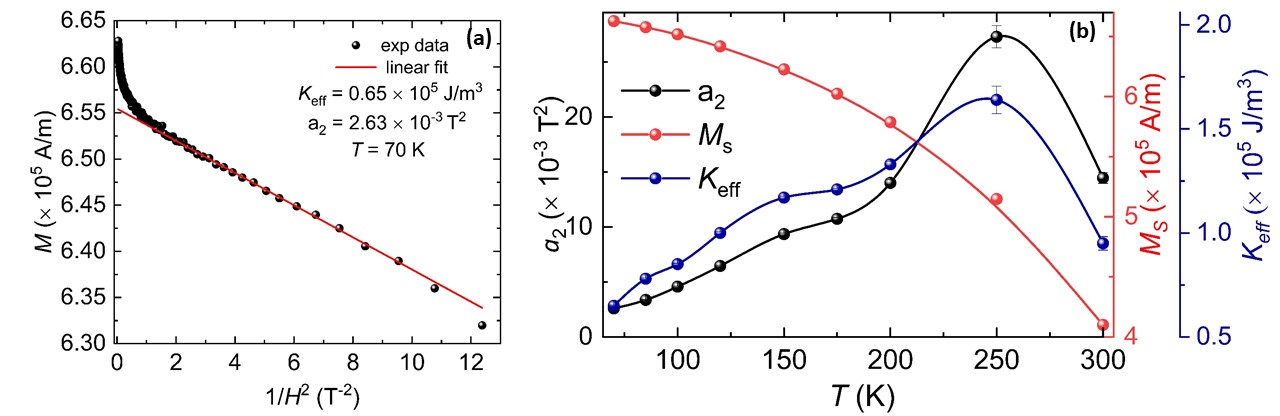}
	\caption{\label{S10} Effective magnetocrystalline anisotropy calculation for Mn$_4$Ga$_2$Sn. (a) Straight line fitting between the magnetization and 1/$H^2$ at 70 K. (b) The fitting parameter a$_2$, saturation magnetization ($M_S$), and calculated anisotropy with varying temperature.}	
\end{figure}

We have calculated effective uniaxial magnetocrystalline anisotropy ($K_{eff}$) of our polycrystalline sample Mn$_4$Ga$_2$Sn applying the law of approach to saturation \cite{Anisotropy}.  The magnetization ($M$) near the saturation ($M$/$M_S$ $\ge$ 0.9), where $M_S$ is saturation magnetization, can be written as $M$ = $M_S$ (1 – a$_2$/$H^2$) with a$_2$ = 4$K^2_{eff}$/15$M_s^2$ is the fitting parameter. From the straight line fitting between $M$ and $H^{-2}$ near the saturation, we have calculated a$_2$ to estimate the possible $K_{eff}$ values. One of the fittings is shown in Figure~\ref{S10}(a). The temperature variation of a$_2$, $M_s$ and $K_{eff}$ are plotted in Figure~\ref{S10}(b). The maximum effective uniaxial anisotropy is observed around 250~K. 
We have also estimated the out-of-plane easy axis anisotropy ($K_u^\perp$) of our sample using the domain wall width, $\delta$= $\pi$ $\sqrt{A/ K_u^\perp}$. In addition, the exchange stiffness constant ($A$) is calculated using the Curie temperature ($T_C$), $A$ $\equiv$ $K_B$$T_C$/$a$, where $K_B$ is Boltzmann constant and $a$ is the lattice parameter of the sample Mn$_4$Ga$_2$Sn. The estimated value of $A$ appears 5$\times$10$^{-12}$~J/m. The average domain wall width of our sample is observed around 22~nm at 250~K. The estimated value of $K_u^\perp$ is 1.02 $\times$ 10$^5$~J/m3. The order of the estimated value of $K_u^\perp$ is also nearly matches with the calculated values of Keff calculated using $M$ vs $H$ data of polycrystalline sample for 250~K, as shown in Figure~\ref{S10}(b). From the 100~K LTEM data we have found  the increase in domain wall width of 29~nm and decrease in calculated $K_u^\perp$ value to 0.58 $\times$ 10$^5$~J/m3.  These findings suggest the decrease in anisotropy of our sample with decreasing temperature.

\subsection{6. Topological Hall effect}

\begin{figure}[h]
	\includegraphics[angle=0,width=16cm,clip]{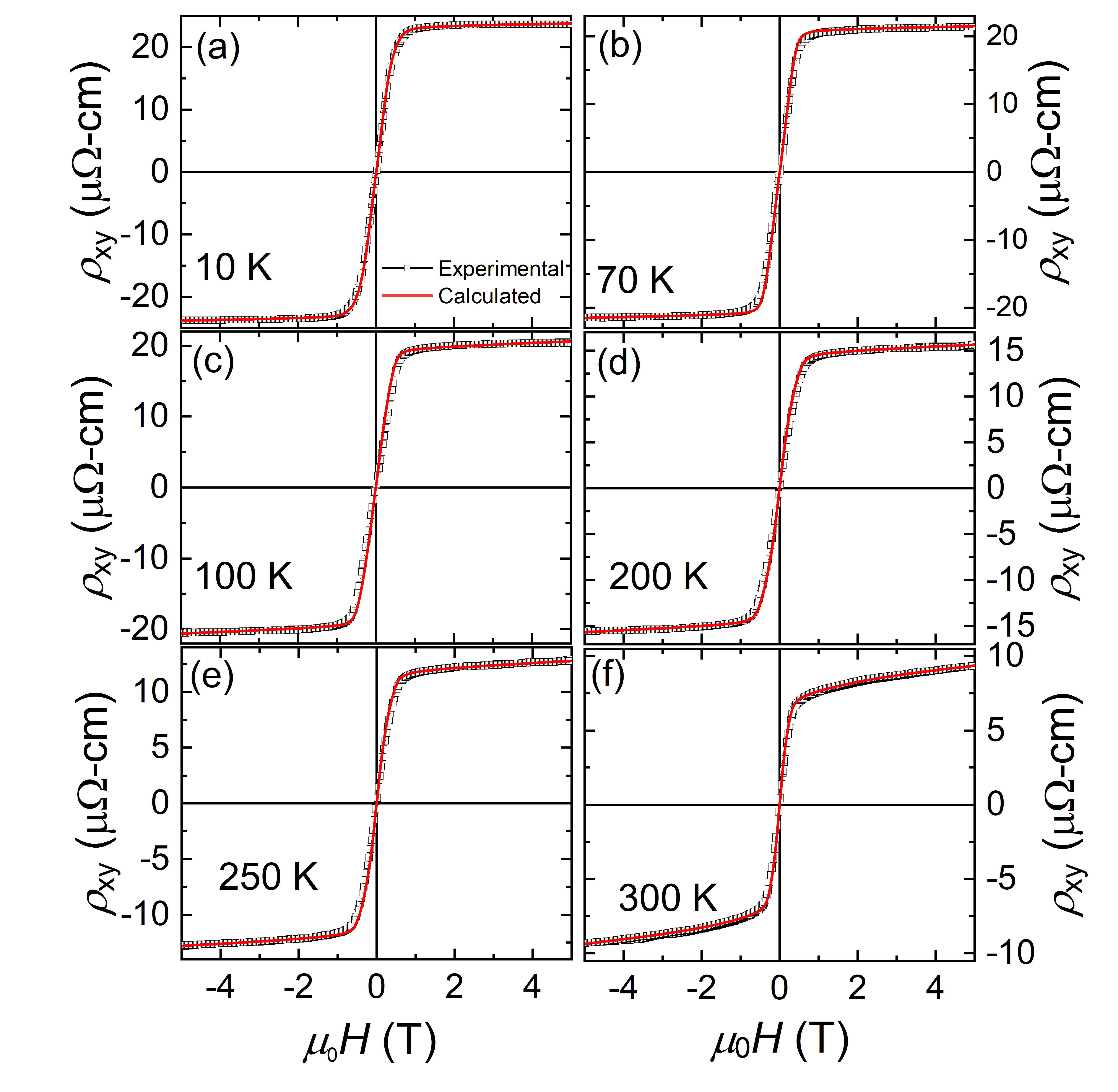}
	\caption{\label{S19} Magnetic field dependent experimental and calculated Hall resistivity for Mn$_4$Ga$_2$Sn polycrystalline sample at different temperatures. Black and red lines represent the experimental and calculated data, respectively.}	
\end{figure}

\begin{figure}[h]
	\includegraphics[angle=0,width=16cm,clip]{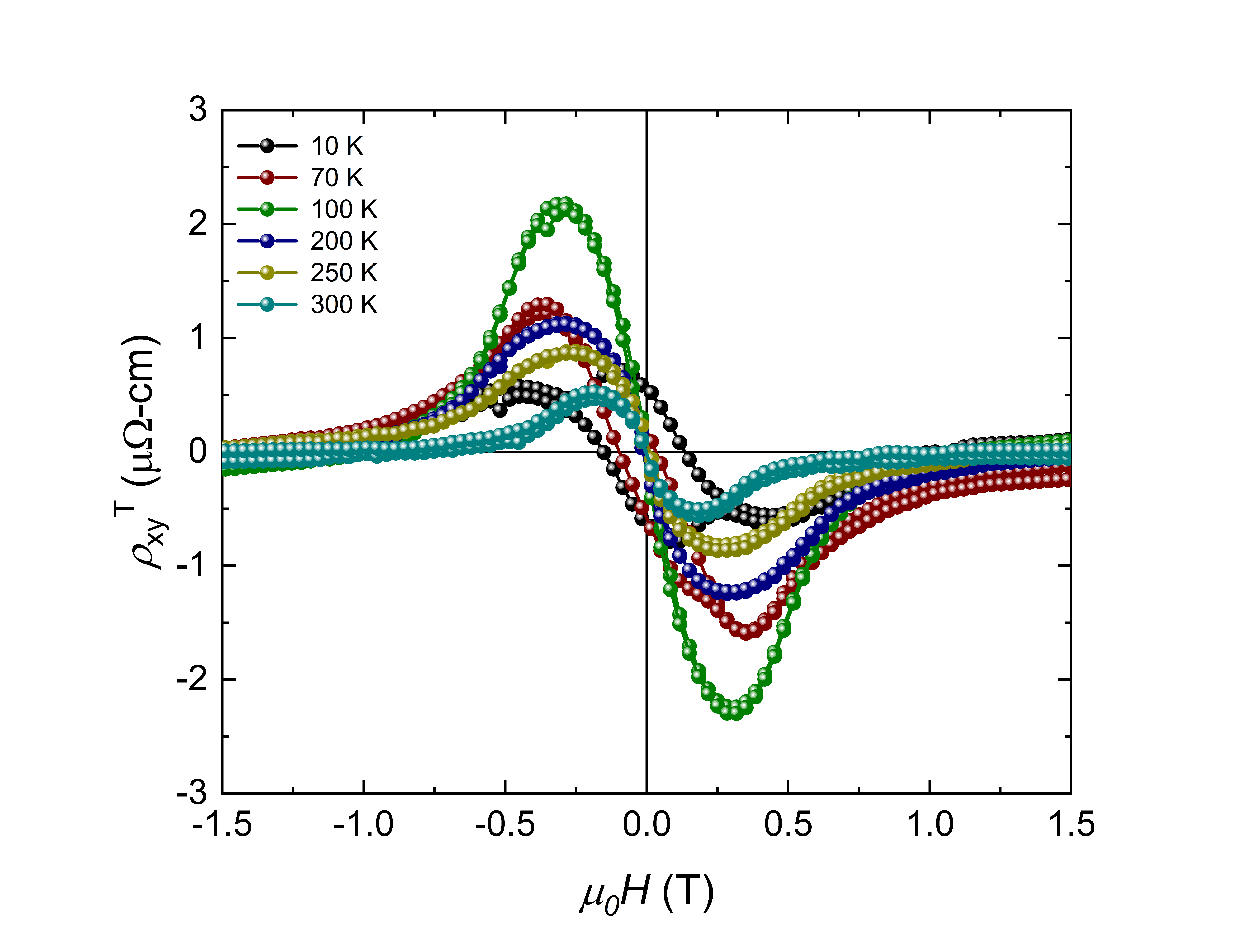}
	\caption{\label{S20} Magnetic field dependent topological Hall resistivity at different temperatures for Mn$_4$Ga$_2$Sn polycrystalline sample.}	
\end{figure}

In case of ferromagnetic material total Hall resistivity can be written as sum of three different contributions like, Normal Hall resistivity ($\rho_{xy}^N$), anomalous Hall resistivity ($\rho_{xy}^A$) and topological Hall resistivity ($\rho_{xy}^T$). So, we can write total Hall effect as, $\rho_{xy}$=  R$_0$$H$ + R$_s$$M$ +  $\rho_{xy}^T$ , where R$_0$ is the normal Hall coefficient, R$_s$ is the anomalous Hall coefficient.

As the longitudinal conductivity of our system belongs to the moderate conductivity regime (10$^4$ $<$ $\sigma_{xx}$ $<$ 10$^6$), the anomalous Hall resistivity in general varies as $\rho_{xx}^2$. So the previous expression can be rewritten as, $\rho_{xy}$ = R$_0$$H$ + b$\rho_{xx}^2$ $M$ + $\rho_{xy}^T$. At very high field the sample transforms into the field polarized state where no topological Hall signal is expected. Hence, at high field we can write the above expression as, $\rho_{xy}$= R$_0$$H$ + b$\rho_{xx}^2$ $M$ or $\rho_{xy}$/H= R$_0$ + b$\rho_{xx}^2$$M/H$. For the extraction of  R$_0$ and b, we have used the straight line fitting of $\rho_{xx}^2$M/H vs. $\rho_{xy}$/H data at high field ($>$ 2 T). Using the extracted b and R$_0$, we have calculated the Hall resistivity from 5 T to -5 T as shown with the red lines in Fig.~\ref{S19}. Then by subtracting the calculated data from the experimental Hall resistivity, we extract the topological Hall resistivity as shown in Fig.~\ref{S20}. As it can be found the maximum THE is visible at 100 K, which is the spin reorientation transition of the present sample. This indicate the major source of the observed THE is the canted magnetic state in the system. The non-vanishing scalar spin chirality of the canted magnetic state can give rise to a large topological Hall signal. Since the size of the skyrmions in the present materials are about 100 nm, the strength of the THE origination from the skyrmion phase must be in the order of few nano-Ohm centimetre, that can be buried inside the THE coming from the spin canting.  

\end{document}